\documentclass[preprint,10pt,sort&compress]{elsarticle}

\usepackage[utf8]{inputenc}
\usepackage{geometry}
\usepackage{mathtools}
\usepackage{amsfonts}
\usepackage{dsfont}
\usepackage{bbm}
\usepackage[normalem]{ulem}
\usepackage{slashed}
\usepackage{tensor}
\usepackage[most]{tcolorbox}
\DeclareMathAlphabet{\mathbbold}{U}{bbold}{m}{n}

\usepackage{pifont}
\newcommand{\cmark}{\ding{51}}
\newcommand{\xmark}{\ding{55}}

\usepackage{graphicx}
\usepackage{array}[=2016-10-06]

\usepackage{placeins}
\usepackage{makecell}
\usepackage{float}

\usepackage{xspace}
\usepackage{xfrac}
\usepackage{hyperref}
\usepackage[nameinlink]{cleveref}

\usepackage[]{tabularx}

\makeatletter
\AddToHook{cmd/appendix/before}{\def\cref@section@alias{appendix}\def\cref@subsection@alias{appendix}}
\makeatother

\usepackage{appendix}
\usepackage{units}

\usepackage[hyphenation]{impnattypo}
\usepackage[all]{nowidow}
\usepackage{microtype}

\usepackage{xifthen}
\usepackage{booktabs}
\usepackage{multirow}
\usepackage[dvipsnames]{xcolor}
\hypersetup{
	colorlinks,
	linkcolor={red!75!black},
	citecolor={blue!75!black},
	urlcolor={blue!75!black},
	pdftitle={FunKit: A computer algebra toolkit for functional approaches},
	pdfauthor={Franz R. Sattler},
}
\usepackage{enumerate}

\usepackage{tikz}
\usetikzlibrary{shapes,arrows,positioning}
\tikzstyle{block} = [rectangle, draw, fill=white!20, 
	text width=10em, text centered, rounded corners, minimum height=2em]
\tikzstyle{dblock} = [draw=black!20!white, text width=10em, dash pattern=on 1pt off 4pt on 6pt off 4pt,
            rectangle, rounded corners, minimum height=2em, text centered]

\newcommand{\wideleft}[1]{\makebox[\linewidth][l]{\hspace{3.0em}#1}}

\usepackage{mmacells}
\mmaDefineMathReplacement[≤]{<=}{\leq}
\mmaDefineMathReplacement[≥]{>=}{\geq}
\mmaDefineMathReplacement[≠]{!=}{\neq}
\mmaDefineMathReplacement{∉}{\notin}
\mmaDefineMathReplacement{∞}{\infty}
\mmaDefineMathReplacement{𝕕}{\mathbbm{d}}
\mmaSet{
	leftmargin=3.0em,
	morefv={gobble=0},
	linklocaluri=mma/symbol/definition:#1,
	morecellgraphics={yoffset=0ex},
	fv={formatcom*=\small\normalfont\ttfamily},
	labelsep=0.em,
}

\usepackage{listings}
\definecolor{backcolor}{rgb}{0.99,0.98,0.98}
\definecolor{string-color}{rgb}{0.3333, 0.5254, 0.345}
\definecolor{darkgrey}{rgb}{0.0627, 0.07, 0.082}
\definecolor{darkred}{rgb}{0.3, 0.05, 0.05}
\definecolor{codeblue}{rgb}{0.2,0.35,0.75}
\definecolor{codepurple}{rgb}{0.38,0.1,0.52}
\definecolor{codegray}{rgb}{0.5,0.5,0.5}
\definecolor{codegreen}{rgb}{0.05,0.3,0.05}
\definecolor{codered}{rgb}{0.6,0.2,0.1}
\definecolor{backgroundColour}{rgb}{0.99,0.99,0.98}
\lstdefinestyle{myStyle}{
	language = C++,
	basicstyle = {\ttfamily \small \color{darkgrey}},
	backgroundcolor = {\color{backcolor}},
	commentstyle=\color{codegreen},
	stringstyle = {\color{string-color}},
	keywordstyle = {\color{codeblue}},
	keywordstyle = [2]{\color{codepurple}},
	keywordstyle = [3]{\color{codered}},
	keywordstyle = [4]{\color{codegray}},
	keywordstyle = [5]{\color{codegreen}},
	otherkeywords = {<, >, :, ::, DiFfRG, constexpr, uint, size_t, &, get, vector, array, Tensor, Scalar, FunctionND},
	morekeywords = [2]{AbstractModel, FEFunctionDescriptor, VariableDescriptor, ExtractorDescriptor, ComponentDescriptor
		TimeStepperSUNDIALS_IDA, UMFPack, Point, real, AD, NoJacobians, FE_AD,LLFFlux,FlowBoundaries
	},
	morekeywords = [3]{DiFfRG, CG, DG, dDG, LDG, def, Variables, dealii, std, autodiff},
	morekeywords = [4]{<, >, :, ::, ;, &},
	morekeywords = [5]{},
	breakatwhitespace=false,
	breaklines=true,
	captionpos=b,
	keepspaces=true,
	numbers=none,
	numbersep=5pt,
	numberstyle=\scriptsize\color{darkred},
	showspaces=false,
	showstringspaces=false,
	showtabs=false,
	tabsize=2
}
\lstdefinestyle{genStyle}{
	basicstyle = {\ttfamily \small \color{darkgrey}},
	backgroundcolor = {\color{backcolor}},
	commentstyle=\color{codegreen},
	stringstyle = {\color{string-color}},
	keywordstyle = {\color{codeblue}},
	breakatwhitespace=false,
	breaklines=true,
	captionpos=b,
	keepspaces=true,
	numbers=none,
	numbersep=5pt,
	numberstyle=\scriptsize\color{darkred},
	showspaces=false,
	showstringspaces=false,
	showtabs=false,
	tabsize=2
}
\lstdefinelanguage{CMake}{%
	morekeywords={if, else, endif, project, cmake_minimum_required, set, find_package, add_executable},%
	sensitive=false,%
	morecomment=[l]{\#},%
	morecomment=[s]{/*}{*/},%
	morestring=[b]",%
	otherkeywords={add_flows, setup_application},%
	keywordstyle = [2]{\color{codepurple}},
	morekeywords = [2]{REQUIRED, HINTS, VERSION, SYSTEM},
}
\lstdefinelanguage{Bash}{%
	morekeywords={if, else, fi, mkdir, cd, cmake, bash, git, make},%
	sensitive=false,%
	morecomment=[l]{\#},%
	morecomment=[s]{/*}{*/},%
	morestring=[b]",%
	keywordstyle = {\color{codeblue}},
	keywordstyle = [2]{\color{codepurple}},
	morekeywords = [2]{$, /, ..},
}

\newcommand{\cpp}{\lstinline[language=C++,style=myStyle]}

\newcommand{\mathem}{\mmaInlineCell{Code}}
\newcommand{\bash}{\lstinline[language=Bash]}

\newcommand{\FunKit}{\texttt{FunKit}\xspace}
\newcommand{\FEDeriK}{\texttt{FEDeriK}\xspace}
\newcommand{\DiRK}{\texttt{DiRK}\xspace}
\newcommand{\DiANE}{\texttt{DiANE}\xspace}
\newcommand{\TRACY}{\texttt{TRACY}\xspace}

\newcommand{\AnSEL}{\texttt{AnSEL}\xspace}
\newcommand{\COEN}{\texttt{COEN}\xspace}

\newcommand{\TensorBases}{\texttt{TensorBases}\xspace}
\newcommand{\FormTracer}{\texttt{FormTracer}\xspace}
\newcommand{\QMeS}{\texttt{QMeS}\xspace}
\newcommand{\DoFun}{\texttt{DoFun}\xspace}
\newcommand{\xAct}{\texttt{xAct}\xspace}
\newcommand{\FORM}{\texttt{FORM}\xspace}
\newcommand{\DiFfRG}{\texttt{DiFfRG}\xspace}

\definecolor{pastelblue}{rgb}{0.4,0.5,0.9}
\definecolor{pastelgreen}{rgb}{0.4,0.9,0.5}

\newcommand{\gettitle}{FunKit: A computer algebra toolkit for functional approaches}
\newcommand{\orcid}[1]{\href{https://orcid.org/#1}{\includegraphics[height=1.9ex,width=1.9ex]{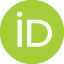}}}

\begin{document}

\begin{frontmatter}
	\renewcommand{\thefootnote}{\fnsymbol{footnote}}
	\title{\gettitle}
	\author[a]{Franz R. Sattler \orcid{0000-0003-1744-9456}\,\footnote{
			\href{mailto:fsattler@physik.uni-bielefeld.de}{fsattler@physik.uni-bielefeld.de}
	}}
	\address[a]{Fakult{\"a}t f{\"u}r Physik, Universit{\"a}t Bielefeld, D-33615 Bielefeld, Germany}
	
	\begin{abstract}
	We introduce \FunKit, a Mathematica package for the derivation and tracing of functional equations from arbitrary master equations.
	\FunKit provides an expression vocabulary and a set of rules that allow for derivations in any given field theory and master equation. It also allows users to add extensions for more specific equation systems.
	Therefore, it can be used in a wide range of situations, for example Dyson--Schwinger or functional RG equations, flowing reparametrisations, nPI equations, (modified) STIs and WTIs, functional Polchinski and Wegner flows, functional master equations with sources, and many others.
	Besides interfacing with the \FORM language to trace large tensor expressions efficiently, \FunKit also provides facilities to export arbitrary Mathematica expressions to C++, Julia or Fortran code, including the results of derivations, which can then be evaluated numerically.
	Both the tracing and code generation can also be used independently and in combination with other packages.
	\end{abstract}

\end{frontmatter}

%%%%%%%%%%%%%%%%%%%
\vspace{-0.2cm}
{\bf PROGRAM SUMMARY}\\\vspace{-0.3cm}

\begin{small}
	\noindent
	{\em Program Title:} \FunKit \\
	{\em Developer's repository link}:  \url{https://github.com/satfra/FunKit} \\
	{\em Licensing provisions:} GPLv3                                   \\
	{\em Programming language:} Mathematica                                  \\
	{\em Computer:}  Linux, macOS, partially Windows                           \\
	{\em Nature of problem:} Derive functional equations in QFTs with any given master equation, from specification of theory and master equation, to executable code.\\
	{\em Solution method:} Full pipeline for typical tasks in computer algebra for functional methods. Implementation of a vocabulary that can handle a wide range of functional expressions including Grassmann fields and functional derivatives. Use of \FormTracer \cite{Cyrol:2016zqb} and \TensorBases \cite{Braun:2025gvq} to allow for tracing of group structures of resulting diagrams. \\
	{\em Unusual features:} Fully user-extensible vocabulary: arbitrary master equations, custom indexed objects, and custom functional-derivative rules. Multi-index derivatives, multi-loop momentum and group routing, source fields, and topological identification of any diagrams. Cached \FORM-based tracing and register-budgeted C++/Julia/Fortran code generation suitable for CPU and GPU solvers. Format-conversion layers to and from \DoFun and \QMeS.	\\
\end{small}

\vspace{-0.7cm}

\tableofcontents

%\tableofcontents
%\vspace{-6ex}
%%%%%%%%%%%%%%%%%%%%%%%%%%%%%%
\section{Introduction}
\label{sec:Introduction}

Functional computations, utilising e.g. the functional Renormalisation Group (fRG) or Dyson--Schwinger equations (DSE), allow for the direct calculation of non-perturbative aspects of quantum field theories. They have been utilised with great success in a broad range of physical situations from condensed matter to QCD to ultracold atoms. For some reviews and introductions to the topic see, e.g., \cite{Roberts:1994dr,Roberts:2000aa,Berges:2000ew,Maris:2003vk,Pawlowski:2005xe,Gies:2006wv,Fischer:2006ub,Delamotte:2007pf,Rosten:2010vm,Metzner_2012,Bashir:2012fs,Fischer:2018sdj,Dupuis:2020fhh,Pawlowski:2020qer,Fu:2022gou,Fischer:2026uni}.

Such functional approaches require the derivation of closed expressions for a set of vertices of a given theory. Expressions for all vertices can be obtained by taking functional derivatives of a given master equation, for example the Wetterich equation \cite{Wetterich:1992yh} and DSEs \cite{Dyson1, Schwinger1, Schwinger2}, or other functional equations derived from symmetry properties.
Such an approach allows non-perturbative access to all correlation functions of the theory while maintaining a finite number of loops in the exact expressions for each vertex. However, an infinite tower of functional equations ensues, which has to be truncated while maintaining all relevant physics.

Many physical systems require sophisticated truncations to reach even qualitative reliability. Thus, the resulting diagrammatic expressions quickly grow very large, and computer-algebraic treatments thereof become a necessity.
For QCD or even just Yang--Mills theory, one obtains flow equations or DSEs which cannot be reliably calculated by hand even in modest truncations.
Additionally, new approaches to functional methods, e.g. spectral fRG flows \cite{Braun:2022mgx} or flowing reparametrisations in fRG, Polchinski or Wegner flows \cite{Ihssen:2023nqd, Ihssen:2024ihp}, require different diagrammatic building blocks that cannot be easily achieved with already established software for automated analytic functional derivations.

There are already a number of powerful libraries available for the computer-algebraic derivation of functional equations. Among these are \DoFun \cite{Alkofer:2008nt,Huber:2011qr,Huber:2019dkb} and \QMeS \cite{Pawlowski:2021tkk}, which concern themselves with the derivation of functional equations from master equations but are restricted to specific master equations. 
The \FormTracer \cite{Cyrol:2016zqb} and the \TensorBases packages \cite{Braun:2025gvq} provide tools to trace the tensorial structure of the resulting diagrams using the symbolic language of \FORM~\cite{Vermaseren:2000nd,Ruijl:2017dtg}.
For a general guide to the derivation of functional equations, also using the aforementioned tools, we refer the reader to the recent review \cite{Huber:2025cbd}.

With \FunKit, we aim to provide a fast, convenient and comprehensive solution for the derivation of arbitrary functional equations.
We provide a complete pipeline to allow for the specification of the theory, the taking of functional derivatives, diagram simplification, the specification of interaction bases, tracing of tensor structures, and the export to C++, Julia or Fortran code for numerical evaluation.

To trace tensor structures, \FunKit builds on \texttt{FormTracer} and \texttt{TensorBases}, while functional derivatives and diagram comparisons are implemented in a reliable manner by using a superindex notation, making it possible to handle any given (user-specified) master equation.

\FunKit also aims to be fast by automatically parallelising tasks during derivations, using a mixture of optimisation algorithms of the \FORM package and Mathematica simplifications to process the output from the tracing of tensor structures, and providing a sophisticated optimisation algorithm for C++, Julia and Fortran code output.
These latter two functionalities, i.e., the helpers for efficient tensor tracing and the generation of highly optimised C++, Julia or Fortran code, can be used independently of the analytic tools. 
This also allows \DoFun or \QMeS output, or otherwise derived equations, to be further processed by \FunKit. We explicitly provide compatibility layers to translate from and to the \DoFun and \QMeS Mathematica notations (see the function references in \Cref{app:all_func}). In particular, this allows for an easy inclusion of \FunKit into already existing computer algebra pipelines.
Furthermore, its modular architecture and high test coverage also make \FunKit  easily maintainable and expandable in the future.

\begin{figure*}[t]
	\centering
	\begin{tikzpicture}[
		node distance=3.5cm,
		block/.style={
			rectangle,
			draw=pastelblue,
			fill=pastelblue!5!white,
			text width=3cm,
			text centered,
			rounded corners,
			minimum height=1.2cm
		},
		arrow/.style={
			draw,
			-stealth,
			thick,
			shorten >=2pt,
			shorten <=2pt
		},
		label/.style={
			text width=2cm,
			text centered,
			font=\small\itshape
		}
		]
		\node[block] (theory) {\textbf{Theory\,\,\,\, Specification}\\
			\textit{Fields, truncation, interaction bases}};
		\node[block, right of=theory] (derivatives) {\textbf{Functional Derivatives}\\
			\textit{FTakeDerivatives[...]}};
		\node[block, right of=derivatives] (truncation) {\textbf{Truncation}\\
			\textit{FTruncate[...]}};
		\node[block, right of=truncation] (routing) {\textbf{Momentum/Index Routing}\\
			\textit{FRoute[...]}};
		\node[block, right of=routing] (tracing) {\textbf{Tensor Tracing}\\
			\textit{FormTrace[...]}};
		\node[block, below of=routing, yshift=1cm] (code) {\textbf{Code Generation}\\
			\textit{Export to C++/Julia/Fortran}};
		\draw[arrow] (theory) -- (derivatives);
		\draw[arrow] (derivatives) -- (truncation);
		\draw[arrow] (truncation) -- (routing);
		\draw[arrow] (routing) -- (tracing);
		\draw[arrow] (tracing) -- (code);
		\node[label, above of=theory, yshift=-2.5cm] {Input};
		\node[label, above of=derivatives, yshift=-2.5cm] {Derivation};
		\node[label, above of=truncation, yshift=-2.5cm] {Selection};
		\node[label, above of=routing, yshift=-2.5cm] {Routing};
		\node[label, above of=tracing, yshift=-2.5cm] {Evaluation};
		\node[label, above of=code, yshift=-2.5cm] {Output};
	\end{tikzpicture}
	\caption{FunKit workflow: From theory specification to numerical evaluation.}
	\label{fig:workflow}
\end{figure*}
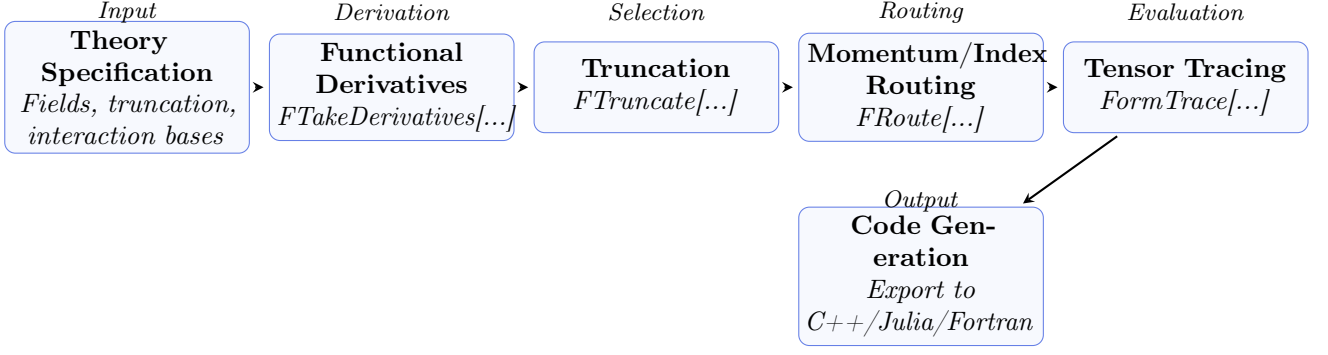

To see what \FunKit does, take the derivation of the flow equation of a two-point function in a scalar~theory, which usually includes the following steps:
\begin{enumerate}
	\setlength\itemsep{0.02ex}
	\item Write down all fields used and the vertex truncation of the theory.
	\item Take two functional derivatives of the Wetterich equation with respect to two scalar fields.
	\item Discard any diagrams that include vertices not in the truncation.
	\item Decide how to propagate momenta through your diagrams and identify the loops in your result.
\end{enumerate}
With \FunKit, these steps translate to certain calls to \FunKit's functions that do this for you:
\begin{enumerate}
	\setlength\itemsep{0.02ex}
	\item Define a \textit{setup} that specifies your field space and truncation.
	\item Use the function \mathem{\mmaDef{FTakeDerivatives}} to take derivatives with respect to two fields.
	\item Use \mathem{\mmaDef{FTruncate}} to discard any diagrams that include vertices not in the truncation.
	\item Use \mathem{\mmaDef{FRoute}} to route all momenta in the expression and group them by loop order.
	\item[(5.)] Create diagrammatic rules and trace group tensors with \mathem{\mmaDef{FMakeDiagrammaticRules}} and \mathem{\mmaDef{FormTrace}}.
	\item[(6.)] Output the result as fully optimised Julia, C++ or Fortran code.
\end{enumerate}
This workflow is also shown in \Cref{fig:workflow} and precisely corresponds to these six steps. Explicitly, to perform the first four steps for the two-point function in a scalar theory one has to call the following with \FunKit:
\begin{mmaCell}{Input}
\mmaDef{Get}["FunKit\`{}"]
\mmaDef{FSetGlobalSetup}[<|"FieldSpace" -> <|"Commuting" -> \{Phi[p]\}|>,
                  "Truncation" -> <|\mmaDef{GammaN} -> \{\{Phi,Phi,Phi\}, \{Phi,Phi,Phi,Phi\}\},
                                    \mmaDef{Propagator} -> \{\{Phi,Phi\}\},
                                    \mmaDef{Rdot} -> \{\{Phi,Phi\}\}|>|>];
\mmaDef{FAddTexStyles}[Phi->"{\textbackslash\textbackslash}phi"];
\mmaDef{FTakeDerivatives}[\mmaDef{WetterichEquation},\{Phi[i1],Phi[i2]\}]//\mmaDef{FTruncate}//\mmaDef{FPlot}//\mmaDef{FRoute}//\mmaDef{FPrint};
\end{mmaCell}
\vspace{-0.5ex}
\noindent\hspace{3.0em}\includegraphics[width=0.4\linewidth]{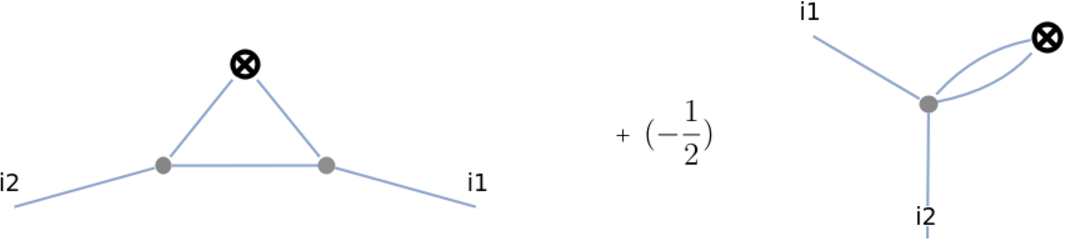}
\vspace{-1.5ex}
\begin{equation*}
\wideleft{\footnotesize$%
	\begin{aligned}\  &\int_{l_1}G_{\phi^{a}\phi^{a}}(l_1,-l_1)\,\Gamma_{\phi^{c}\phi^{a}\phi^{d}}(l_1-p_1,-l_1,p_1)\,G_{\phi^{c}\phi^{c}}(l_1-p_1,p_1-l_1)\,\Gamma_{\phi^{a}\phi^{c}\phi^{d}}(l_1,p_1-l_1,-p_1)\,G_{\phi^{a}\phi^{a}}(l_1,-l_1)\,\partial_t R_{\phi^{a}\phi^{a}}(-l_1,l_1)
		\\ &\,+\,\int_{l_1}\left(-\frac{1}{2}\,G_{\phi^{a}\phi^{a}}(l_1,-l_1)\,\Gamma_{\phi^{a}\phi^{a}\phi^{c}\phi^{c}}(l_1,-l_1,-p_1,p_1)\,G_{\phi^{a}\phi^{a}}(l_1,-l_1)\,\partial_t R_{\phi^{a}\phi^{a}}(-l_1,l_1)\right)
	\end{aligned}
$}
\end{equation*}
The above code showcases already the overall workflow of \FunKit, which will be described in greater detail within the following sections.

The current work is structured in the following way:
\Cref{sec:Notation} briefly introduces the notation used throughout the paper, for more detailed explanations, see \Cref{app:notation_details}.
Then, we introduce the basic syntax for describing functional expressions in \Cref{sec:FEDeriK} and explain how to specify a theory, truncation and how to momentum- and index-route diagrams.
In \Cref{sec:TRACY}, the specification of tensor bases and diagrammatic expressions is explained. Therein, we also show how to trace the resulting diagrams.
\Cref{sec:COEN} briefly shows how to generate computer code in different languages from these equations.
With \Cref{sec:YMExample}, we showcase a full example using \FunKit, with the full source code provided, from the derivation of Yang--Mills theory fRG flows, until the numerical evaluation.
Finally, in \Cref{sec:comparisons}, we benchmark and compare \FunKit against other frameworks for computer algebra in functional contexts.

We add green boxes throughout the text that provide additional information on the algorithms used by \FunKit. In blue boxes, we provide quick reference overviews for the reader. Also, note that in \Cref{app:all_func} a full reference is provided, where all functions provided by \FunKit are listed and explained.

\begin{table*}[t]
	\begin{tcolorbox}[title=Available example notebooks,
		title filled=false,
		colback=pastelblue!5!white,
		colframe=pastelblue,
		fonttitle=\bfseries]
		\small
		\begin{tabular}{p{6.0cm} p{9.2cm}}
			\toprule
			\textbf{Notebook (repo path)} & \textbf{Description} \\
			\midrule
			\texttt{examples/FunKitPaper.nb} & Notebook that reproduces the code examples used throughout this paper; intended as a guided, runnable companion to the manuscript. \\[1ex]
			\texttt{examples/mSTI-Yang-Mills.nb} & Example showing the calculation of the modified Slavnov--Taylor identity (mSTI) for the gluon two-point function. \\[1ex]
			\texttt{examples/Yang-Mills.nb} & Derivation examples for Yang–Mills theory: DSEs and flow equations with the truncation used in the paper. \\[1ex]
			\texttt{examples/Yang-Mills/} & Derivation of a closed flow system for Yang--Mills theory, together with a working numerical implementation of the generated~code. \\[1ex]
			\texttt{examples/ScalarTheory.nb} & Minimal scalar-field examples: DSEs and flow equations. \\[1ex]
			\texttt{examples/Yukawa.nb} & Example setup and derivations for a Yukawa model: DSEs and flow equations.  \\[1ex]
			\texttt{examples/FlowingReparametrisation.nb} & Example setup and derivations for fRG equations in a Yukawa model with flowing reparametrisation.\\
			\texttt{examples/CompositeOperators.nb} & Derivation of correlation functions of a two-fermion composite.
			%\bottomrule
		\end{tabular}
	\end{tcolorbox}
	\caption{Overview of example notebooks included in the \FunKit repository.}
	\label{tab:examples}
\end{table*}

%%%%%%%%%%%%%%%%%%%%%%
\subsection*{Installation \& Examples}

To follow the code examples, one may install the package from inside a Mathematica notebook using
\begin{mmaCell}{Input}
\mmaDef{Import}["https://raw.githubusercontent.com/satfra/FunKit/main/FunKitInstaller.m"]
\end{mmaCell}
If the package is already installed, import \FunKit with
\begin{mmaCell}{Input}
\mmaDef{Get}["FunKit\`{}"]
\end{mmaCell}
\FunKit's source code and examples are available in a GitHub repository \cite{FunKitRepo}.
The \FunKit package has no requirements except for either a Mathematica version $\geq 11.0$ or a corresponding free WolframEngine installation.
While Linux and macOS are fully supported, \FunKit on Windows has only partial feature support, as \FORM does not support execution on Windows systems. Tracing group structures and creating diagrammatic rules is thus unavailable by default on Windows. If the user is able to provide a working \FormTracer installation, they can set \mathem{\mmaDef{$UseFORMOnWindows}=True} before loading the package to enable all functionality.
We recommend using \FunKit via WSL (Windows Subsystem for Linux), which always allows one to utilise the full feature~set.

A Notebook containing all code written below can be found at \texttt{examples/FunKitPaper.nb} in the repository \cite{FunKitRepo}.
Besides the guided explanations in this work, documentation on all functions provided by \FunKit can be accessed through their usage strings, i.e.,
\begin{mmaCell}{Input}
\mmaDef{FDOp}::\mmaCmt{usage}
\end{mmaCell}
\begin{mmaCell}{Output}
FDOp[field[index]]
Represents a functional derivative operator \({\delta}\)/\({\delta}\)field acting on everything to its right. 
FDOp operators are resolved using FResolveDerivatives or FResolveFDOp functions.
\end{mmaCell}
Inside a notebook, a brief overview can be accessed using \mathem{\mmaDef{FInfo[]}}.
We also provide several example notebooks where the derivation of certain theories with several functional master equations is shown, which we list in~\Cref{tab:examples}.

\begin{table*}[t]
	\begin{tcolorbox}[title=Quick Notation Reference,
		title filled=false,
		colback=pastelblue!5!white,
		colframe=pastelblue,
		fonttitle=\bfseries]
		\begin{tabular}{p{3.8cm} p{2.2cm}  p{8.7cm}}
			\toprule
			Mathematica&Expression&Description \\
			\midrule
			\mathem{\mmaDef{AnyField}[a]} & $\Phi^a$ &
			\small Superfield with multi-index $a$. \\[2ex]
			\mathem{Psi[a]} &  $\psi^a$ &
			\small Explicit field with a multi-index $a$, e.g. for a fermion $a=(d,F,C)$ (Dirac, flavour and colour groups). \\[2ex]
			$\gamma$\mathem{[{...},{a,b}]} & $\gamma^{ab}$ &
			\small Field metric used to raise/lower superindices: $\gamma^{ab}\Gamma_b=\Gamma^a$, see \eqref{eq:metric}. \\[2.5ex]
			\mathem{\mmaDef{FMinus}[{...},{a,b}]} & $(-1)^{ab}$ &
			\small Sign object for fermion permutations: $\hspace{4.2cm}$ for example \mathem{\mmaDef{FMinus}[{Psi,Psi},{a,b}]} = $-1$, see \eqref{eq:fminus}. \\[2ex]
			\mathem{\mmaDef{FEx}[\mmaDef{FTerm}[...], ...]} & $ $ &
			\small \mathem{\mmaDef{FTerm}}: a non‑commutative product (ordered factors). $\hspace{1.6cm}$ \mathem{\mmaDef{FEx}}: sum of \mathem{\mmaDef{FTerm}}s (main container). \\[2ex]
			\mathem{\mmaDef{GammaN}}, \mathem{\mmaDef{S}}, \mathem{\mmaDef{Propagator}}, $\quad$ \mathem{\mmaDef{R}}, \mathem{\mmaDef{Rdot}} & $\Gamma$, $S$, $G$, $\qquad\quad$ $R$, $\partial_tR$ &
			\small Indexed objects: 1PI vertices $\Gamma$ (\mathem{\mmaDef{GammaN}}), classical vertices $S$ (\mathem{\mmaDef{S}}), propagator $G^{ab}$ (\mathem{\mmaDef{Propagator}}), regulator $R_{ab}$ (\mathem{\mmaDef{R}}) and its derivative~$\partial_t R_{ab}$~(\mathem{\mmaDef{Rdot}}). \\[3ex]
			\mathem{\mmaDef{FDOp}[...]} & $\frac{\delta}{\delta \Phi^a}$ &
			\small Functional derivative operator acting on the remainder of the term. E.g. \mathem{\mmaDef{FDOp}[A[i]]} = $\frac{\delta}{\delta A^i}$, or \mathem{\mmaDef{FDOp}[\mmaDef{R}[{A,A},{-a,-b}]]} = $\frac{\delta}{\delta R_{ab}}$.\\[2.5ex]
			\mathem{\mmaDef{SymmetryFactor}[...]} & $\textrm{Sym}_{a_1a_2\ldots}$ &
			The symmetry factor $\textrm{Sym}_{a_1a_2\ldots}$ that arises in \labelcref{eq:symDeriv} is represented as \mathem{\mmaDef{SymmetryFactor}[{\mmaDef{AnyField},\mmaDef{AnyField},...},{a1,a2,...}]}.\\
			%
			%\bottomrule
			%
		\end{tabular}
	\end{tcolorbox}
	\caption{Brief overview for the most important symbols used in the notation of {\FunKit}, as introduced step by step in~\Cref{sec:FEDeriK}.}
	\label{tab:notation}
\end{table*}

%%%%%%%%%%%%%%%%%%%%
\section{Preliminaries \& Notation}
\label{sec:Notation}

%%%%%%%%%%%%%%%%%%%%%%%%%%%%%%
%\subsubsection{Superfield notation}

To enable a reliable and concise handling of indices, in particular fermionic ones, \FunKit makes abundant use of the superfield notation used in \cite{Pawlowski:2005xe}. Importantly, keeping track of signs for Grassmann fields is ensured by making use of such a notation. In the following, we briefly state the most important definitions; for detailed derivations we refer the reader to \Cref{app:notation_details} and \cite{Pawlowski:2021tkk}.

In order to have a concise notation for any given QFT, we use a single superfield $\Phi$, whose (upper) multi-index $\Phi^a$ contains field type, momentum and group indices, e.g.\ $a = (A,p, (\mu, c))$ and therefore $\Phi^a = A_\mu^c(p)$.

An Einstein summation convention is adopted throughout this work, i.e., a sum is performed over any index pair where one index is raised and one lowered. As an example, in $\Gamma_{ab}G^{bd}$, a sum is performed over the index~$b$. This implies a sum over all field types, all associated group indices, and an integration over all momenta that occur. So, given a theory with gluons and ghosts $\bar{c}$, $c$, the resulting index expansion looks like
\begin{align*}
	\Gamma_{ab}G^{bd} = \sum_{\mu e}\int_p\Gamma_{a\,A^e_{\mu}(p)}G^{A^e_{\mu}(p)\,d} + \sum_{e}\int_p\Gamma_{a\,c^e(p)}G^{c^e(p)\,d} + \sum_{e}\int_p\Gamma_{a\,\bar{c}^e(p)}G^{\bar{c}^e(p)\,d}\,.
\end{align*}

Anti-commutation of Grassmann-valued fields is encoded in a field metric $\gamma^{ab}$ \cite{Pawlowski:2005xe, Braun:2025gvq},
\begin{equation}\label{eq:metric}
	\gamma_{ab} = \gamma^{ab} = \begin{cases}
		\begin{pmatrix}
			0 & -1 \\
			1 & 0
		\end{pmatrix}\,\delta_{ab}
		& \text{if $a$ and $b$ are fermionic,} \\[3ex]
		\delta_{ab} & \text{if $a$ and $b$ are bosonic,}   \\[2ex]
		0               & \text{otherwise}
		\,.
	\end{cases}
\end{equation}
For the purpose of the metric \labelcref{eq:metric}, we group together fermions and anti-fermions. Explicitly, if $\psi$, $\bar{\psi}$ are, respectively, a fermion and its anti-fermion, we have
\begin{align}
	\gamma^{\psi\psi} = 0\,,\quad
	\gamma^{\psi\bar\psi}=-1\,,\quad
	\gamma^{\bar\psi\psi}=1\,,\quad
	\gamma^{\bar\psi\bar\psi} = 0\,.
\end{align}
The metric can be used to lower and raise indices of arbitrary tensors involving fields. We use the northwest-southeast convention, i.e., raise indices to the northwest, lower indices to the southeast:
\begin{equation}
	\Phi^a = \gamma^{ab}\Phi_b \,,
	\quad \quad
	\Phi_a = \Phi^b\gamma_{ba} \,.
\end{equation}
We also introduce the fermionic sign shorthand,
\begin{equation}
	(-1)^{ab} = \begin{cases}
		-1 & \text{if both $a$ and $b$ are fermionic,} \\[1ex]
		1 & \text{otherwise}
		\,.
	\end{cases}
	\label{eq:fminus}
\end{equation}
From the quantum equation of motion one derives the relation between the connected propagator $G^{ab}$ and the 1PI two-point function $\Gamma_{ab}$, namely $G^{bc}\Gamma_{ca} = \gamma^b_{\phantom{b}a}$, as well as the derivative rule
\begin{align}\label{eq:derivG}
	\left(\frac{\delta}{\delta \Phi^f}G^{ba}\right)
	&= (-1)(-1)^{bf}(-1)^{dd}\, G^{bc}\Gamma_{cfd}G^{da}\,.
\end{align}
For a derivation, see \Cref{app:notation_details}. Multi-index functional derivatives, relevant for nPI equations, are also defined in \Cref{app:notation_details}.
For explicit tensors, we work in Euclidean space throughout this work, but \FunKit does not restrict one to that.\\
In the following sections, all expressions used above will be introduced as symbols in Mathematica step by step. A brief translation table for the most important symbols can be found below in \Cref{tab:notation}.

%%%%%%%%%%%%%%%%%%%%%%%%%%%%%%
\subsection{Sign conventions}
\label{sec:Signs}

Sign conventions are a common source of confusion in computer algebra systems involving Grassmann fields. In \FunKit, and by extension also in \TensorBases, which is relevant for \Cref{sec:TRACY}, we follow consistently the following notation:
\begin{enumerate}
	\item Subscripts on correlation functions or other functionals represent derivatives, performed from the right to the left. Explicitly, this means that
	\begin{align}
		F_{a_1\ldots a_n} = \frac{\delta }{\delta \Phi^{a_1}} \cdots \frac{\delta }{\delta \Phi^{a_n}} F[\Phi]\,,
	\end{align}
	where $F$ stands in for any functional.
	
	\item \FunKit always orders fields such that commuting fields come first, then anti-Grassmann fields, then Grassmann fields. This can be also changed by the user, if so desired, see \Cref{app:all_func}.
\end{enumerate}
The conventions chosen here match exactly the ones of \QMeS. \DoFun adopts a slightly different convention in the ordering of the fields, see \Cref{app:notation_details}.

For later reference, let us here consider an example: we take the ghost part of the 1PI generating function $\Gamma$ of a gauge-fixed Yang--Mills theory, explicitly
\begin{align}
	\Gamma[\Phi] &= \int_x \bar{c}_a \left(-\partial_\mu D^{ab}_\mu \right) c_b \,+\,\ldots
	\\[1ex]
	 &= \int_p \bar{c}_a(-p) p^2 \delta^{ab} c_b(p) + i\int_{p,q}\bar{c}_a(p) \left( p_\mu gf^{abc}A^c_\mu(-p-q)\right) c_b(q)\,,
\end{align}
where $a,b,c$ are adjoint colour indices, $f^{abc}$ is the structure constant of the Lie algebra, and $\int_p = \int \frac{d^dp}{(2\pi)^d}$.
Note also that we have performed a partial integration in the second term.
Consequently, we immediately infer the ghost two-point function and the ghost-gluon vertex as
\begin{alignat}{2}
	\Gamma_{\bar{c}_ac_b}(p,q) &=
		\frac{\delta}{\delta\bar{c}_a(p)}\frac{\delta}{\delta c_b(q)} \Gamma[\Phi] &&=
		-p^2\delta_{ab}
			\delta(p+q)
	\,,
	\notag\\[1ex] 
	\Gamma_{A^c\bar{c}_a c_b}(r,p,q) &=
		\frac{\delta}{\delta A^c_\mu(r)}\frac{\delta}{\delta\bar{c}_a(p)}\frac{\delta}{\delta c_b(q)} \Gamma[\Phi] &&=
	 	-i p_\mu g\, f^{abc} 
			\delta(r+p+q)
	\,.	
\end{alignat}
%

%%%%%%%%%%%%%%%%%%%%%%%%%%%%%%
\section{Deriving functional equations}
\label{sec:FEDeriK}

%%%%%%%%%%%%%%%%%%%%%%%%%%%%%%
\subsection{General usage}
\label{sec:FEDeriK_notation}

%%%%%%%%%%%%%%%%%%%%%%%%%%%%%%
\subsubsection{Syntax}
\label{sec:FEDeriK_syntax}

We use the Wetterich equation \cite{Wetterich:1992yh} as a first example of how to write down a functional equation in \FunKit. It~reads
\begin{equation}
	\partial_t\Gamma=\frac{1}{2}G^{ab}\partial_tR_{ab}\,,
	\label{eq:Wetterich}
\end{equation}
with the 1PI generating functional $\Gamma$, the connected two-point function (propagator) $G^{ab}$, and the regulator $R_{ab}$, which suppresses infrared modes of the given QFT. All modes below a certain scale $k$ are suppressed, and the RG-time $t$ is defined as $t=\ln(k/\Lambda)$, where $\Lambda$ is the initial cutoff scale of the flow.
In \FunKit, \labelcref{eq:Wetterich} can be written as the \mathem{\mmaDef{FTerm}}

\begin{mmaCell}{Input}
wEq = \mmaDef{FTerm}[1/2, \mmaDef{Propagator}[\{\mmaDef{AnyField},\mmaDef{AnyField}\},\{a,b\}], \mmaDef{Rdot}[\{\mmaDef{AnyField},\mmaDef{AnyField}\},\{-a,-b\}]]
\end{mmaCell}
A single term in a functional equation is described using the \mathem{\mmaDef{FTerm}} tag, which encloses a list of factors, multiplied in the precise order as they appear inside \mathem{\mmaDef{FTerm}}. In other words, it represents a non-commutative product.
Furthermore, we have used \FunKit's common notation for objects with indices. Any indexed object takes two lists, where the first contains all involved fields and the second contains the corresponding (super-) indices. 
For example, with this notation we can represent
\begin{align}
	 G^{A^a\,b}\simeq\texttt{\small Propagator[\{\textcolor{blue}{A},AnyField\},\{a,b\}]}\,.
\end{align}
Here, the \mathem{\mmaDef{AnyField}} tag represents an unspecified field, which can be later on expanded into explicit fields, while \mathem{A} is an explicit field.%
\footnote{An alternative notation, activated with \texttt{FSetNotationB[]}, pairs fields with indices as \texttt{obj[f1[i1],f2[i2],...]}, similar to how \DoFun's notation works. All operations work identically in both notations.}

\FunKit knows by default the following objects:
\begin{mmaCell}{Input}
\mmaDef{FShowObjects}[]
\end{mmaCell}
\begin{mmaCell}{Output}
Field
FMinus
GammaN
Phidot
Propagator
R
Rdot
S
SymmetryFactor
\({\gamma}\)
\end{mmaCell}
Lowered superindices are, in both notations, indicated by a minus sign; upper indices carry no sign. This notation is adapted from \xAct \cite{xact2025}. As an example, the correlation function $\Gamma_{a\,b\,A^c}^{\phantom{a\,b\,A^c}\bar{c}^d}$, with the fields of $a$ and $b$ undetermined, may be expressed~as
\begin{mmaCell}{Input}
\mmaDef{GammaN}[\{\mmaDef{AnyField},\mmaDef{AnyField},A,cb\},\{-a,-b,-c,d\}]
\end{mmaCell}

Sums of functional equations are represented using the \mathem{\mmaDef{FEx}} tag. \mathem{\mmaDef{FEx}} can only contain \mathem{\mmaDef{FTerm}}s, and the \mathem{\mmaDef{FEx}} represents their sum.
As an example, we take the generalised flow equation
\begin{equation}
	\partial_t \Gamma = -\dot\Phi^a\Gamma_a + \frac{1}{2}G^{ac}\left(\gamma_c^{\phantom{c}b}\partial_t + 2\frac{\delta\dot{\Phi}^b}{\delta\Phi^c}\right){R}_{ab}
	\,,
\end{equation}
which has three distinct terms. With \FunKit's notation, it can be expressed~as
\begin{mmaCell}{Input}
\mmaDef{FAddCorrelationFunction}[PhiDot];
\mmaDef{FSetUnorderedIndices}[PhiDot,1];
\mmaDef{FEx}[
  \mmaDef{FTerm}[-PhiDot[\{\mmaDef{AnyField}\},\{a\}], \mmaDef{GammaN}[\{\mmaDef{AnyField}\},\{-a\}]],
  \mmaDef{FTerm}[1/2, \mmaDef{Propagator}[\{\mmaDef{AnyField},\mmaDef{AnyField}\},\{a,c\}],
    \mmaDef{FEx}[
      \mmaDef{FTerm}[\({\gamma}\)[\{\mmaDef{AnyField},\mmaDef{AnyField}\},\{-c,b\}], \mmaDef{Rdot}[\{\mmaDef{AnyField},\mmaDef{AnyField}\},\{-a,-b\}]],
      \mmaDef{FTerm}[\mmaDef{FDOp}[\mmaDef{AnyField}[c]],PhiDot[b],\mmaDef{R}[\{\mmaDef{AnyField},\mmaDef{AnyField}\},\{-a,-b\}]]
     ]
   ]
 ]
\end{mmaCell}
\begin{mmaCell}{Output}
FEx[FTerm[-PhiDot[\{AnyField\},\{a\}],GammaN[\{AnyField\},\{-a\}]],
  FTerm[1/2,Propagator[\{AnyField,AnyField\},\{a,c\}],\({\gamma}\)[\{AnyField,AnyField\},\{-c,b\}],Rdot[\{AnyField,AnyField\},\{-a,-b\}]],
  FTerm[1/2,Propagator[\{AnyField,AnyField\},\{a,c\}],FDOp[AnyField[c]],PhiDot[b],R[\{AnyField,AnyField\},\{-a,-b\}]]]
\end{mmaCell}
In the above, we have first informed \FunKit about a user-defined correlation function, \mathem{\mmaDef{PhiDot}} $\equiv \dot{\Phi}$, and instructed it to never change the position of the last index (i.e., the superindex specifying $\Phi$ itself).
Furthermore, we have already used \mathem{\mmaDef{FDOp}}, which represents a functional derivative operator, and nested \mathem{\mmaDef{FEx}} and \mathem{\mmaDef{FTerm}}. As visible from the output, \FunKit automatically expanded this into three \mathem{\mmaDef{FTerm}} expressions.

We remark that \FunKit knows by default the correlation function \mathem{\mmaDef{Phidot}}, which is precisely defined this way, but for demonstration purposes we have re-defined it here as \mathem{\mmaDef{PhiDot}}.
\FunKit will now automatically treat \mathem{\mmaDef{PhiDot}} as a field-dependent function, and take into account derivatives thereof.

\mathem{\mmaDef{FEx}} and \mathem{\mmaDef{FTerm}} have some internal rules to automatically move numeric values to the first factor in any \mathem{\mmaDef{FTerm}} and expand any appearance of sums to the enclosing \mathem{\mmaDef{FEx}}. This behaviour can be seen, e.g., in
\begin{mmaCell}{Input}
\mmaDef{FEx}[\mmaDef{FTerm}[a+b,2,4c,\mmaDef{FEx}[\mmaDef{FTerm}[d+e]]]]
\end{mmaCell}
\begin{mmaCell}{Output}
\mmaDef{FEx}[\mmaDef{FTerm}[8, a + b, c, d + e]]
\end{mmaCell}
\mathem{\mmaDef{FEx}} and \mathem{\mmaDef{FTerm}} can be multiplied using non-commutative multiplication, 
\begin{mmaCell}{Input}
\mmaDef{FEx}[\mmaDef{FTerm}[a+b,2]]**\mmaDef{FEx}[\mmaDef{FTerm}[d+e]]
\end{mmaCell}
\begin{mmaCell}{Output}
\mmaDef{FEx}[\mmaDef{FTerm}[2, a + b, d + e]]
\end{mmaCell}
Normal multiplication immediately yields an error, as the factors in a given \mathem{\mmaDef{FTerm}} are not necessarily commuting.
\begin{mmaCell}{Input}
\mmaDef{FEx}[\mmaDef{FTerm}[a+b,2]]*\mmaDef{FEx}[\mmaDef{FTerm}[d+e]]
\end{mmaCell}
\begin{mmaCell}{Output}
\$Aborted
\end{mmaCell}
\FunKit is rather conservative in how much it expands sums in a given expression, as in actual diagrammatic expressions with tensor bases fully inserted this can lead to significant performance degradation in Mathematica.

%%%%%%%%%%%%%%%%%%%%%
\subsubsection{Field setup}

Most functions defined by \FunKit require the specification of a \textit{setup}. A setup is given by an \mathem{\mmaDef{Association}} with the mandatory key \mathem{"FieldSpace"} and further optional keys, which we will specify below.

The field content of the theory is specified using another \mathem{\mmaDef{Association}}, usually with two entries, \mathem{"Commuting"}, and \mathem{"Grassmann"}. The value of these entries is a list of all fields occurring in the theory. For example, the fields of a gauge-fixed Yang--Mills theory can be given as
\begin{mmaCell}{Input}
fields = <|
  "Commuting" -> \{A[p,\{v,c\}]\},
  "Grassmann" -> \{\{cb[p,\{c\}], c[p,\{c\}]\}\}
 |>;
\end{mmaCell}
The gauge field \mathem{A} is a commuting field and has no partner field. In contrast, there are anti-ghosts and ghosts \mathem{cb}, \mathem{c}, hence they are specified as a pair in a nested list.
Fields are always specified together with their full index structure. 
In the above case of Yang--Mills theory, all fields have both momentum and additional group structure.  The momentum is then the first argument of the field and a list of all group indices gives the (optional) second argument.
A momentum variable is always obligatory, i.e., a purely scalar field space would be specified as
\begin{mmaCell}{Input}
fieldsPhi = <|"Commuting" -> \{\(\pmb{\phi}\)[p]\}|>;
\end{mmaCell}
The resulting setup can be then defined as
\begin{mmaCell}{Input}
setup = <|"FieldSpace" -> \mmaDef{fields}|>;
\mmaDef{FSetGlobalSetup}[setup];
\end{mmaCell}
All functions of \FunKit support two ways to call them. Either a specific setup is passed as the first argument of the function, and the function uses only the local information of the given setup. Or, a \textit{global setup} is specified using \mathem{\mmaDef{FSetGlobalSetup}}. Then, all functions of \FunKit can be called without explicitly passing a setup, and they simply use the setup that has been globally specified by the user. To shorten the presentation, in this work we exclusively use this second mode of operation.

Also, useful for deriving purely abstract expressions without explicit fields, \FunKit provides the pre-made \mathem{\mmaDef{FEmptySetup}}, which contains nothing but an empty field space.

%%%%%%%%%%%%%%%%%%%%
\subsubsection{Source terms}

Source terms are treated as non‑dynamical external fields in \FunKit: they are not expanded in truncation operations, but participate normally in all other operations. Furthermore, they are always reordered to be in the right-most position when appearing in derivatives of correlators.
This is convenient for derivations that involve external sources, for example the Slavnov--Taylor identities (STIs) in gauge theories, where sources appear but should not be promoted to dynamical degrees of freedom.
An example Yang--Mills setup with explicit source entries is:
\begin{mmaCell}{Input}
fieldsYMmSTI = <|
  "Commuting" -> \{A[p,\{v,c\}]\},
  "Grassmann" -> \{\{cb[p,\{c\}], c[p,\{c\}]\}\},
  "CommutingSource" -> \{Qcb[p], Qc[p]\},
  "GrassmannSource" -> \{QA[p,\{v,c\}]\}
 |>;
\end{mmaCell}
In the field space definition, Grassmann sources do not need to be paired with anti-fields.
We provide the full calculation of the modified STI for the gluon two-point function as an example in an fRG situation within the notebook \texttt{examples/mSTI-Yang-Mills.nb}.

%%%%%%%%%%%%%%%%%%%%%%%%%%%%%%%%%%%%%%%%
\subsection{Deriving diagrams}

With the \mathem{\mmaDef{setup}} defined, one can immediately start deriving diagrams.
Above, we have defined the Wetterich equation as \mathem{\mmaDef{wEq}}. We can obtain the RG-flow of the gluon two-point function using
\begin{mmaCell}{Input}
\mmaDef{FTakeDerivatives}[\mmaDef{wEq}, \{A[i1], A[i2]\}];
\end{mmaCell}
This will yield a rather large expression, which we can visualise using the \mathem{\mmaDef{FPrint}} and \mathem{\mmaDef{FPlot}} commands, which will be explained in more detail within \Cref{sec:DiANE}.
\begin{mmaCell}{Input}
\mmaDef{FTakeDerivatives}[\mmaDef{wEq}, \{A[i1], A[i2]\}]//\mmaDef{FPrint}//\mmaDef{FPlot};
\end{mmaCell}
\vspace{-1.5ex}
\begin{equation*}
\wideleft{\footnotesize$%
		\begin{aligned}\  &\Big(\,\frac{1}{2} (-1)^{A^{i_1}a} (-1)^{A^{i_2}a} (-1)^{cc} (-1)^{ee}
			\\&\,\,+\frac{1}{2} (-1)^{A^{i_1}f} (-1)^{A^{i_2}f} (-1)^{af} (-1)^{bA^{i_1}} (-1)^{ba} (-1)^{bb} (-1)^{cA^{i_1}} (-1)^{cb} (-1)^{dA^{i_2}} (-1)^{dc} (-1)^{dd} (-1)^{eA^{i_2}} (-1)^{ed} (-1)^{fe}\,\Big)
			\\&\,\quad\times\,G^{ab}\,\Gamma_{bA^{i_1}c}\,G^{cd}\,\Gamma_{dA^{i_2}e}\,G^{ef}\,\partial_t R_{af}
			\\ &\,+\,\left(-\frac{1}{2} (-1)^{A^{i_1}a} (-1)^{A^{i_1}b} (-1)^{A^{i_2}a} (-1)^{cc}\,G^{ab}\,\Gamma_{A^{i_1}bA^{i_2}c}\,G^{cd}\,\partial_t R_{ad}\right)
		\end{aligned}
$}
\end{equation*}
\vspace{-2ex}

\noindent\hspace{3.0em}\includegraphics[width=0.38\linewidth]{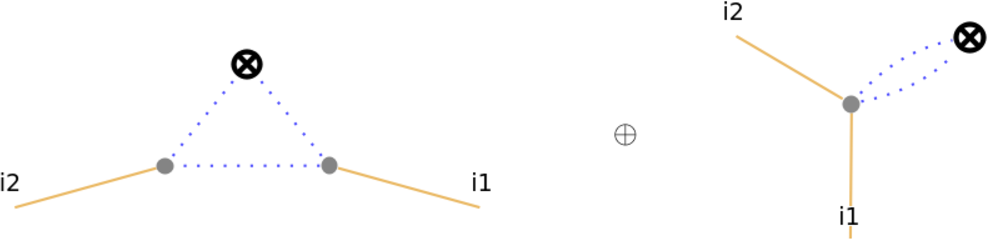}\\
In the above, $(-1)^{ab}$ is the fermionic minus sign as defined in \labelcref{eq:fminus}, represented in \FunKit as \mmaInlineCell{Input}{\mmaDef{FMinus}[\{\mmaDef{AnyField},\mmaDef{AnyField}\},\{a,b\}]}. Note additionally that, $\gamma_{ab}$ is represented as \mmaInlineCell{Input}{\mmaDef{\(\pmb{\gamma}\)}[\{\mmaDef{AnyField},\mmaDef{AnyField}\},\{-a,-b\}]}, see also \Cref{tab:notation} for an overview of the notation.

\FunKit has automatically simplified the diagrams resulting from the above derivation. For finer control, the functions \mathem{\mmaDef{FResolveFDOp}} and \mathem{\mmaDef{FSimplify}} allow resolving derivatives and simplifying expressions step by step. In general, however, we recommend using \mathem{\mmaDef{FTakeDerivatives}}, as it provides the most automated workflow and generates additional simplification hints such as symmetry properties of the derivatives.
We describe the simplification and derivative algorithms used by \FunKit in more detail in the green boxes below.

\begin{tcolorbox}[title=Algorithm details: Diagram simplification,
	title filled=false,
	colback=pastelgreen!5!white,
	colframe=pastelgreen!73!black,
	fonttitle=\bfseries]
	
	With \mathem{\mmaDef{FSimplify}} we introduce an algorithm for the simplification of large sums of connected diagrams. \mathem{\mmaDef{FSimplify}} can detect identical diagrams and performs the following steps to find all matches and sum~them:
	\begin{enumerate}
		\setlength\itemsep{0.1ex}
		\item All Diagrams with exactly the same content of objects and the same set of open indices are grouped together.
		\item Each group is then processed in parallel: Pairs of diagrams are sequentially built and compared.
		\item The comparison of two diagrams is performed by traversing the graphs of both at the same time and checking for identity. If the diagrams are disconnected, their sub-graphs are compared pair-wise. 
		\item If symmetries of open indices are provided, all permutations (with possible prefactors, e.g. minus signs for fermionic permutations) are also explored.
		\item If the graphs are identical, they are merged. When merging, the second diagram is prepended with a factor containing all signs needed to give it precisely the same index ordering and structure as the first diagram.
	\end{enumerate}
	
\end{tcolorbox}

\begin{tcolorbox}[title=Algorithm details: Derivatives,
	title filled=false,
	colback=pastelgreen!5!white,
	colframe=pastelgreen!73!black,
	fonttitle=\bfseries]
	
	\mathem{\mmaDef{FTakeDerivatives}} performs the following steps:
	\begin{enumerate}
		\setlength\itemsep{0.1ex}
		\item The derivative list is used to construct a symmetry list that encodes all permutations of the open superindices that give the same result. This is automatically annotated to the result later.
		\item Starting with the right-most field in the derivative list, \mathem{\mmaDef{FDOp}}s are attached to the left of each \mathem{\mmaDef{FTerm}} in the given expression.
		\item The expression is given to \mathem{\mmaDef{FResolveDerivatives}}, which recursively invokes \mathem{\mmaDef{FResolveFDOp}}:
		\begin{enumerate}
			\setlength\itemsep{0.05ex}
			\item Identify the right-most \mathem{\mmaDef{FDOp}} in the given \mathem{\mmaDef{FTerm}}.
			\item Realise the product rule: If there are $n$ factors in the given \mathem{\mmaDef{FTerm}} following the position of the \mathem{\mmaDef{FDOp}}, construct $n$ new terms $F_1,\ldots,F_n$. In the term $F_i$, the \mathem{\mmaDef{FDOp}} has been commuted with $i-1$ terms that follow, attaching the resulting sign from commutation to $F_i$.
			\item If \mathem{\mmaDef{FSetAutoSimplify}} is toggled on (default) and the number of generated terms is large, use \mathem{\mmaDef{FSimplify}} to reduce the result.
			\item If any of the $F_i$ contain another \mathem{\mmaDef{FDOp}}, start from (a) with this term.
		\end{enumerate}
		\item Collect all new \mathem{\mmaDef{FTerm}} in a \mathem{\mmaDef{FEx}}.
		\item If \mathem{\mmaDef{FSetAutoSimplify}} is toggled on (default), use \mathem{\mmaDef{FSimplify}} to reduce the result.
	\end{enumerate}
	If the number and size of the expressions is large, \mathem{\mmaDef{FResolveFDOp}} automatically chooses to work in parallel over the generated terms.
	
\end{tcolorbox}

%%%%%%%%%%%%%%%%%%%%%%%%%%%%%%%%%%%%%%%%
\subsection{Truncating the output}

The above equation still contains undetermined fields, which are expressed by superindices without fields in the equations and by dotted blue lines in the diagrams.
To insert explicit fields, we first have to supply a truncation. This is done by creating a list for every correlation function or indexed object which contains all field combinations to be taken into account:
\begin{mmaCell}{Input}
truncation = <|
  \mmaDef{GammaN} -> \{\{A,A\},\{A,A,A\},\{A,A,A,A\},\{A,cb,c\},\{cb, c\}\},
  \mmaDef{Propagator} -> \{\{A,A\},\{cb, c\}\},
  \mmaDef{Rdot} -> \{\{A,A\},\{cb, c\}\},
  \mmaDef{S} -> \{\{A,A\},\{A,A,A\},\{A,A,A,A\},\{cb,c\},\{cb,c,A\}\},
  \mmaDef{Field} -> \{\{\}\}
 |>;
\end{mmaCell}
The truncation for \mathem{\mmaDef{Field}} is left empty and \FunKit will set all field values to zero when truncating. If $A$ has a finite expectation value, as in finite-temperature Yang--Mills theory, one would change this to \mathem{\mmaDef{Field} -> {{A}}}. In that case, only explicit ghost and anti-ghost fields are set to zero.
Conversely, omitting a key from the truncation table altogether keeps every instance of that object unchanged: \FunKit only filters objects whose head appears as a key.

Next, we modify the setup as
\begin{mmaCell}{Input}
\mmaDef{setup} = <|"FieldSpace" -> \mmaDef{fields}, "Truncation" -> \mmaDef{truncation}|>;
\mmaDef{FSetGlobalSetup}[\mmaDef{setup}];
\mmaDef{FSetTexStyles}[cb -> "{\textbackslash\textbackslash}bar\{c\}"];
\end{mmaCell}
In the above, we have additionally told \FunKit how to render anti-ghosts when translating them to \LaTeX{} code.

We can now easily truncate the above expression:
\begin{mmaCell}{Input}
\mmaDef{FTakeDerivatives}[\mmaDef{wEq},\{A[i1],A[i2]\}]//\mmaDef{FTruncate}//\mmaDef{FPrint}//\mmaDef{FPlot};
\end{mmaCell}
\vspace{-1.5ex}
\begin{equation*}
\wideleft{\footnotesize$%
	\begin{aligned}\  &\left(-2\,G^{c^{a}\bar{c}^{b}}\,\Gamma_{A^{i_1}\bar{c}^{c}c^{a}}\,G^{c^{d}\bar{c}^{c}}\,\Gamma_{A^{i_2}\bar{c}^{e}c^{d}}\,G^{c^{f}\bar{c}^{e}}\,\partial_t R_{\bar{c}^{b}c^{f}}\right)
		\\ &\,+\,G^{A^{a}A^{b}}\,\Gamma_{A^{c}A^{b}A^{i_1}}\,G^{A^{c}A^{d}}\,\Gamma_{A^{e}A^{d}A^{i_2}}\,G^{A^{e}A^{f}}\,\partial_t R_{A^{f}A^{a}}
		\\ &\,+\,\left(-\frac{1}{2}\,G^{A^{a}A^{b}}\,\Gamma_{A^{c}A^{b}A^{i_2}A^{i_1}}\,G^{A^{c}A^{d}}\,\partial_t R_{A^{d}A^{a}}\right)
	\end{aligned}
$}
\end{equation*}
\vspace{-1ex}

\noindent\hspace{3.0em}\includegraphics[width=0.78\linewidth]{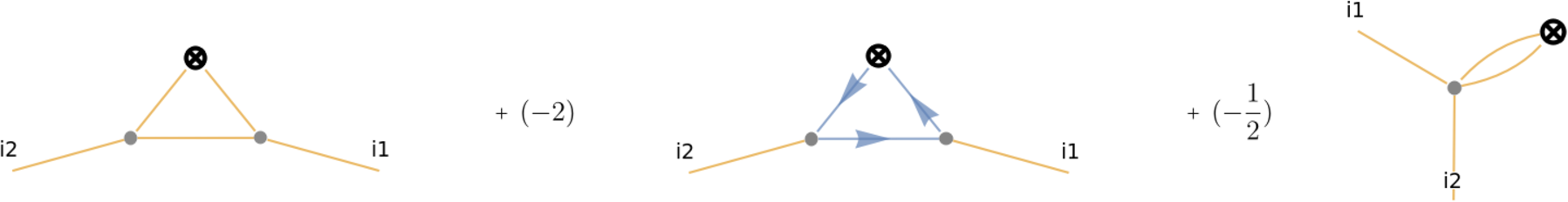}\\
\begin{tcolorbox}[title=Algorithm details: Truncation,
	title filled=false,
	colback=pastelgreen!5!white,
	colframe=pastelgreen!73!black,
	fonttitle=\bfseries]

	\mathem{\mmaDef{FTruncate}} performs the following steps:
	\begin{itemize}
		\item If no undetermined fields \mathem{\mmaDef{AnyField}} are present, apply the truncation table directly and return the result. This zeros all objects whose field content is not listed in the setup's key \mathem{"Truncation"}.
		\item \textbf{Propagator truncation}: For each propagator-like object (\mathem{\mmaDef{Propagator}}, \mathem{\mmaDef{Rdot}}, \mathem{\mmaDef{R}}) with \mathem{\mmaDef{AnyField}}, enumerate its explicit field alternatives from the truncation table. 
		Incrementally replace each unresolved propagator with the sum of its alternatives, propagate the field assignments to connected objects, and fully expand the product over the sum.
		After each expansion step, apply the truncation table. This early pruning prevents combinatorial blowup by discarding invalid branches before subsequent propagators are expanded.
		\item \textbf{Vertex truncation}: For terms containing no propagator-like objects, valid field assignments are instead enumerated by imposing consistency at every closed index incrementally and validating each surviving assignment against the truncation table.
		\item \textbf{Post-processing:} Contract residual metric factors, fix index positions, and reorder fields into canonical form.
		\item If the corresponding flag is set, automatically  invoke \mathem{\mmaDef{FSimplify}} to merge identical surviving diagrams.
	\end{itemize}
\end{tcolorbox}

%%%%%%%%%%%%%%%%%%%%%%
\subsection{Routing momenta and group indices}

Finally, the superfield indices have to be turned into explicit group indices, and the momenta occurring in diagrams need to be routed, thereby identifying the loop 
momenta.
\FunKit uses the information given through the \mathem{"FieldSpace"} definition in the setup to do both momentum and index routing. This can be immediately done using the \mathem{\mmaDef{FRoute}} function:
\begin{mmaCell}{Input}
\mmaDef{FTakeDerivatives}[\mmaDef{wEq}, \{A[i1], A[i2]\}]//\mmaDef{FTruncate}//\mmaDef{FRoute}//\mmaDef{FPrint};
\end{mmaCell}
\vspace{-2.ex}
\begin{equation*}
\wideleft{\footnotesize\medmuskip=0mu$%
	\begin{aligned}\  &\int_{l_1^{\text{(f)}}}\Big(-2\,G_{c^{a}\bar{c}^{b}}(l_1^{\text{(f)}},-l_1^{\text{(f)}})\,\Gamma_{A^{m}\bar{c}^{c}c^{a}}(p_1,l_1^{\text{(f)}}-p_1,-l_1^{\text{(f)}})\,G_{c^{e}\bar{c}^{c}}(l_1^{\text{(f)}}-p_1,p_1-l_1^{\text{(f)}})\,\Gamma_{A^{n}\bar{c}^{g}c^{e}}(-p_1,l_1^{\text{(f)}},p_1-l_1^{\text{(f)}})\,\times
		\\[-1ex]&\hspace{10cm}
		\,G_{c^{i}\bar{c}^{g}}(l_1^{\text{(f)}},-l_1^{\text{(f)}})\,\partial_t R_{\bar{c}^{b}c^{i}}(l_1^{\text{(f)}},-l_1^{\text{(f)}})\Big)
		\\ &\,+\,\int_{l_1}G_{A^{a}A^{b}}(l_1,-l_1)\,\Gamma_{A^{c}A^{a}A^{m}}(l_1-p_1,-l_1,p_1)\,G_{A^{e}A^{c}}(l_1-p_1,p_1-l_1)\,\Gamma_{A^{n}A^{g}A^{e}}(-p_1,l_1,p_1-l_1)\,G_{A^{i}A^{g}}(l_1,-l_1)\,\partial_t R_{A^{i}A^{b}}(-l_1,l_1)
		\\ &\,+\,\int_{l_1}\left(-\frac{1}{2}\,G_{A^{a}A^{b}}(l_1,-l_1)\,\Gamma_{A^{i}A^{c}A^{a}A^{j}}(-p_1,l_1,-l_1,p_1)\,G_{A^{e}A^{c}}(l_1,-l_1)\,\partial_t R_{A^{e}A^{b}}(-l_1,l_1)\right)
	\end{aligned}
$}
\end{equation*}
Note in the above that \mathem{\mmaDef{FRoute}} marks loop momenta that are purely fermionic, here $l_1^{(f)}$. 
This is particularly useful in finite-temperature calculations, where purely fermionic closed loops have to be summed over fermionic Matsubara frequencies. 
The \mathem{\mmaDef{FRoute}} command groups together terms with the same loop order and additionally provides both the explicit indices of all external momenta and the names of all loop momenta.
We explain the algorithm used by the routing in the green box at the end of this section.

As an example, we consider here the DSE for the gluon two-point function. In general, the DSE master equation is given as
\begin{align}
	\Gamma_{c} = \left\langle \frac{\delta S}{\delta \phi^c} \right\rangle_{\phi^a = \Phi^a + G^{ab}\frac{\delta}{\delta\Phi^b}}
	\,,
\end{align}
and it can be readily obtained in \FunKit by using \mathem{\mmaDef{FMakeDSE}[...]}, providing a setup and a field that is inserted for the superindex $c$.
Note that \mathem{\mmaDef{FMakeDSE}} can also be readily constructed by the user: First, $\delta S / \delta \phi^c$ can be directly created from a given classical action, e.g. using \mathem{\mmaDef{FMakeClassicalAction}}. Next, all fields can be replaced with the rule \mathem{f[a_]->\mmaDef{FEx}[\mmaDef{FTerm}[f[a]],\mmaDef{FTerm}[\mmaDef{Propagator}[{f,\mmaDef{AnyField}},{a,b}],\mmaDef{FDOp}[\mmaDef{AnyField}[b]]]]}. Finally, one can resolve all derivative operators with \mathem{\mmaDef{FResolveDerivatives}}. This is precisely what the convenience function \mathem{\mmaDef{FMakeDSE}} internally does.

Now, to obtain an equation for the gluon two-point function, we choose $c = A^a_{\mu}$ and take a further functional derivative with respect to $A$ (note that we don't provide a setup, as a global one has been specified before): 

\begin{mmaCell}{Input}
gluonDSE = \mmaDef{FTakeDerivatives}[\mmaDef{FMakeDSE}[A[i1]], \{A[i2]\}]//\mmaDef{FTruncate}//\mmaDef{FRoute};
{gluonDSE}["0-Loop"]//\mmaDef{FPrint}
{gluonDSE}["1-Loop"]["LoopMomenta"]
{gluonDSE}["1-Loop"]["ExternalIndices"]
{gluonDSE}["1-Loop"]//\mmaDef{FPrint};
{gluonDSE}["2-Loop"]//\mmaDef{FPlot};
\end{mmaCell}
\vspace{-3.0ex}
\begin{equation*}
\wideleft{\footnotesize$S_{A^{a}A^{b}}(-p_1,p_1)$}
\end{equation*}
\vspace{-4.0ex}
\begin{mmaCell}{Output}
<|Expression->FEx[FTerm[S[\{A,A\},\{\{-p1,\{v2,c2\}\},\{p1,\{v1,c1\}\}\}]]],
  ExternalIndices->\{i1->\{p1,\{v1,c1\}\},i2->\{-p1,\{v2,c2\}\}\}, LoopMomenta->\{\}|>
\end{mmaCell}
\begin{mmaCell}{Output}
\{l1|lf1\}
\end{mmaCell}
\begin{mmaCell}{Output}
\{i1->\{p1,\{v1,c1\}\},i2->\{-p1,\{v2,c2\}\}\}
\end{mmaCell}
\vspace{-2.0ex}
\begin{equation*}
\wideleft{\footnotesize\medmuskip=0mu$%
		\begin{aligned}\    &\int_{l_1}\left(-\frac{1}{2}\,S_{A^{a}A^{b}A^{i}}(-l_1-p_1,l_1,p_1)\,G_{A^{c}A^{a}}(-l_1-p_1,l_1+p_1)\,\Gamma_{A^{e}A^{c}A^{j}}(-l_1,l_1+p_1,-p_1)\,G_{A^{b}A^{e}}(-l_1,l_1)\right)
			\\ &\,+\,\int_{l_1}\frac{1}{2}\,S_{A^{a}A^{b}A^{e}A^{f}}(l_1,-l_1,-p_1,p_1)\,G_{A^{b}A^{a}}(l_1,-l_1)
			\\ &\,+\,\int_{l_1^{\text{(f)}}}\left(-S_{A^{i}\bar{c}^{a}c^{b}}(p_1,-l_1^{\text{(f)}}-p_1,l_1^{\text{(f)}})\,G_{c^{c}\bar{c}^{a}}(-l_1^{\text{(f)}}-p_1,l_1^{\text{(f)}}+p_1)\,\Gamma_{A^{j}\bar{c}^{e}c^{c}}(-p_1,-l_1^{\text{(f)}},l_1^{\text{(f)}}+p_1)\,G_{c^{b}\bar{c}^{e}}(-l_1^{\text{(f)}},l_1^{\text{(f)}})\right)
		\end{aligned}
$}
\end{equation*}
\vspace{-1.0ex}
\noindent\hspace{3.0em}\includegraphics[width=0.75\linewidth]{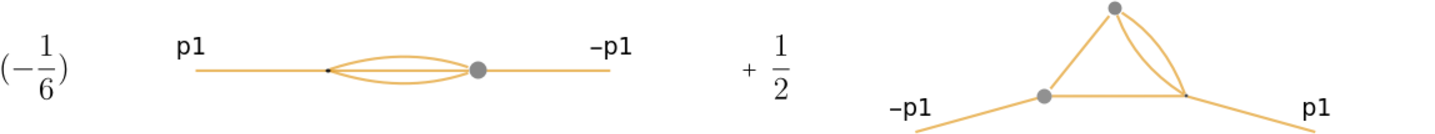}\\[1ex]
The output of \mathem{\mmaDef{FRoute}} returns an association with keys as \mathem{"0-Loop"}, \mathem{"1-Loop"}, and so on, where all diagrams are grouped that have the same number of loops. Each of these elements is again an association that holds all necessary information about the result.
The respective keys are:
\begin{itemize}
	\item \mathem{"Expression"}: The expression itself, which is a list of all diagrams.
	\item \mathem{"ExternalIndices"}: This is a list of all open indices in the expression, with all momentum- and group indices expanded. 
	This is especially helpful to convert projectors from superindex-notation to real indices, as we will make use of below in \Cref{sec:TRACY}.
	\item \mathem{"LoopMomenta"}: This is either a list of all loop momenta, or a list of alternative patterns for loop momenta. If a loop is purely fermionic, it gets assigned a momentum name with an additional \texttt{f}. This information is relevant for finite temperature calculations, but can be usually discarded in the vacuum.
\end{itemize}

In the above example, we also see that the command \mathem{\mmaDef{FPrint}} can be inserted during a chain of operations, since it prints the LaTeX output as a side effect and returns its argument unchanged.

\begin{tcolorbox}[title=Algorithm details: Routing,
	title filled=false,
	colback=pastelgreen!5!white,
	colframe=pastelgreen!73!black,
	fonttitle=\bfseries]
	
	\mathem{\mmaDef{FRoute}} works by performing the following steps:
	\begin{enumerate}
		\item Every open superindex is replaced by a unique set of explicit indices and momenta.
		\item All closed indices are replaced by explicit indices. Closed group indices are automatically contracted, but every superindex pair is assigned a unique loop momentum.
		\item Starting with vertices that have open indices, we iterate over all vertices. Momentum conservation is invoked to replace a loop momentum. The routing is performed such that fermionic momenta are correctly routed through fermionic legs (which is relevant for finite-temperature calculations).
	\end{enumerate}
	Diagrams are then grouped by the number of remaining loop momenta, which naturally gives the number of loops within each diagram.
	
\end{tcolorbox}

%%%%%%%%%%%%%%%%%%%%
\subsection{Teaching FunKit}
\label{sec:FEDeriK_teaching}

Advanced users may be interested in expanding the vocabulary and derivative rules of \FunKit. 
This is explicitly supported to allow for very general applications, and we explain it briefly in the following. 
\subsubsection*{Adding new objects}
Use the \mathem{\mmaDef{FAdd...}} family of helpers to register new symbols so they participate correctly in ordering, truncation and derivative algorithms:
\begin{mmaCell}{Input}
\mmaDef{FAddObject}[MyTag];               (* general object that may carry indices but does not
                                    react to reordering or contraction, acting as a "bystander",
                                    like Sym and FMinus *)
\mmaDef{FAddIndexedObject}[MyIndexed];    (* object that carries field/index lists, must be 
                                    correctly contracted *)
\mmaDef{FAddOrderedObject}[MyOrdered];    (* participates in canonical ordering *)
\mmaDef{FAddCorrelationFunction}[MyCorr]; (* treated as an n-point correlation function *)
\end{mmaCell}
Each helper registers and protects the symbol and adjusts internal lists so the new object behaves like the built‑in ones (e.g. \mathem{\mmaDef{Propagator}}, \mathem{\mmaDef{GammaN}}, \mathem{\mmaDef{S}}, ...).
\subsubsection*{Fixing index positions}
Some indexed objects contain indices that must not be reordered by the field‑ordering routines (for example, the last index of a flowing reparametrisation function). Use
\begin{mmaCell}{Input}
\mmaDef{FSetUnorderedIndices}[PhiDot, 1];
\end{mmaCell}
to mark the last index position of \mathem{PhiDot} as unordered. 
This is relevant insofar as the flowing reparametrisation \mathem{PhiDot} can be chosen differently for each field and thus changing the last index of \mathem{PhiDot} does not mean reordering derivatives, but actually using a different function.
The second argument may be a single integer or a list of positions, counted from the right.
\subsubsection*{Custom functional derivative rules}
When the built‑in derivative rules are insufficient, one may add user rules that the internal function for taking functional derivatives, \mathem{\mmaDef{FunctionalD}}, applies during derivative resolution. Rules are added with
\begin{mmaCell}{Input}
\mmaDef{FAddFDRule}[objectPattern, wrtPattern, resultExpr];
\end{mmaCell}
Examples:
\begin{mmaCell}{Input}
\mmaDef{FAddFDRule}[customFunc[x_], Phi[y_], customResult[y,x]];
\mmaDef{FAddFDRule}[myObj[{A,A},{i_,j_}], A[k_], myRuleResult[{k,i,j}]];
\end{mmaCell}
\mathem{\mmaDef{FAddFDRule}} uses \mathem{\mmaDef{HoldAll}} to keep patterns intact and is applied by \mathem{\mmaDef{FunctionalD}} before the default derivative rules. Provide patterns on the left and the desired evaluated result on the right (use Mathematica patterns where needed).
\subsubsection*{Clearing custom rules}
To remove all user‑added derivative rules and restore the default behaviour call
\begin{mmaCell}{Input}
\mmaDef{FClearFDRules}[];
\end{mmaCell}
This resets the internal list of user rules.

%%%%%%%%%%%%%%%%%%%%%%%%%%%%%%%%%%%%%%%%
\subsection{Equation and Diagram output}
\label{sec:DiANE}

As already demonstrated in the previous subsections, \FunKit provides the commands \mathem{\mmaDef{FPrint}} and \mathem{\mmaDef{FPlot}} for rendering equations in \LaTeX\xspace and plotting Feynman diagrams, respectively. Both commands accept expressions at any stage of the derivation pipeline and return them unchanged, so they can be freely inserted during chained computations. The typesetting code can also be extracted as a string using \mathem{\mmaDef{FTex}}, which is convenient for copying equations into manuscripts. Internally, \FunKit uses \texttt{MaTeX} \cite{szabolcs_horvat_2024_10828124} for \LaTeX\xspace rendering, which requires \bash{pdflatex} to be available on the system.

To adjust the plot styling, one can add another key to the setup, which is \mathem{"DiagramStyling"}. Note that one could also style \mathem{\mmaDef{AnyField}}.
\begin{mmaCell}{Input}
diagramStyling = <|"EdgeStyles" -> \{
  A -> \{Orange\},
  c -> \{Black, \mmaDef{Dashed}\}
  \}|>;
Setup["DiagramStyling"] = diagramStyling;
FSetGlobalSetup[Setup];
\end{mmaCell}
If we now plot again the flow of the gluon propagator, as calculated above, it looks now like this:
\begin{mmaCell}{Input}
\mmaDef{FTakeDerivatives}[\mmaDef{wEq},\{A[i1],A[i2]\}]//\mmaDef{FTruncate}//\mmaDef{FPlot};
\end{mmaCell}
\vspace{-1ex}
\noindent\hspace{3.0em}\includegraphics[width=0.8\linewidth]{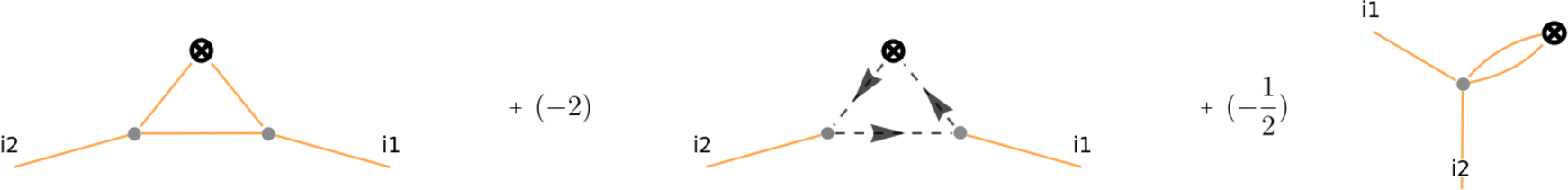}\\
Additionally, one can give hints to \FunKit which objects should be interpreted as edges, and which ones as vertices, by using the keys \mathem{"Vertices"} and \mathem{"Edges"}, and override the default vertex shape and size via \mathem{"VertexStyles"} and \mathem{"VertexSizes"}.
Additionally, external-leg labelling can be turned off with \mathem{"ExternalIndexLabels"->\mmaDef{False}}.
As an example, one could define the following styling options:
\begin{mmaCell}{Input}
\mmaDef{diagramStyling} = <|
  "EdgeStyles" -> \{A -> \{Orange\},
                   c -> \{Black, \mmaDef{Dashed}\}\},
  "Vertices" -> \{MyCounterterm,ExternalSource\},
  "Edges" -> \{MyPropagator\},
  "VertexStyles" -> \{MyCounterterm -> \mmaDef{Graphics}[\{\mmaDef{Red},\mmaDef{Rectangle}[\{-1,-1\},\{1,1\}]\}]\},
  "VertexSizes" -> \{MyCounterterm -> 0.2\},
  "ExternalIndexLabels" -> \mmaDef{False}
 |>;
\end{mmaCell}
This requires of course that objects such as \mathem{MyCounterterm} have been registered with \FunKit before, see \Cref{sec:FEDeriK_teaching}. Entries in \mathem{"VertexStyles"} and \mathem{"VertexSizes"} take precedence over the built-in defaults, so they can also be used to re-skin the canonical heads (\mathem{\mmaDef{GammaN}}, \mathem{\mmaDef{S}}, \mathem{\mmaDef{Field}}, \mathem{\mmaDef{Rdot}}, \mathem{\mmaDef{R}}, \mathem{\mmaDef{Phidot}}).

%%%%%%%%%%%%%%%%%%%%%%%%%%%%%%%%%%%%%%%%
\section{Diagrammatic rules and tracing of expressions}
\label{sec:TRACY}

After having derived the functional equation, it is necessary to use a set of diagrammatic rules to obtain an explicit expression for a given diagram. \FunKit provides functions to automatically create such rules from a chosen set of tensor bases. Nevertheless, we will first present a brief example how to define such rules by hand, and then introduce the automatic rule-creation included in \FunKit.

%%%%%%%%%%%%%%%%%%%%%%%%%%%%%%
\subsection{Manually creating diagrammatic rules}

To see how to create the diagrammatic Feynman rules by hand, consider the ghost propagator flow, given by
\begin{mmaCell}{Input}
diagGhostFlow = \mmaDef{FTakeDerivatives}[\mmaDef{wEq}, \{cb[i1], c[i2]\}]//\mmaDef{FTruncate}//\mmaDef{FPlot}//\mmaDef{FRoute}//\mmaDef{FPrint};
\end{mmaCell}
\vspace{-1.1ex}
\begin{equation*}
	\wideleft{\footnotesize$%
		\begin{aligned}\  &\int_{l_1}\left(-G_{A^{a}A^{b}}(-l_1,l_1)\,\Gamma_{A^{b}\bar{c}^{m}c^{d}}(-l_1,p_1,l_1-p_1)\,G_{c^{d}\bar{c}^{f}}(p_1-l_1,l_1-p_1)\,\Gamma_{A^{g}\bar{c}^{f}c^{n}}(l_1,p_1-l_1,-p_1)\,G_{A^{g}A^{j}}(-l_1,l_1)\,\partial_t R_{A^{j}A^{a}}(-l_1,l_1)\right)
			\\ &\,+\,\int_{l_1}(-G_{c^{a}\bar{c}^{b}}(l_1+p_1,-l_1-p_1)\,\Gamma_{A^{c}\bar{c}^{m}c^{a}}(l_1,p_1,-l_1-p_1)\,G_{A^{c}A^{f}}(-l_1,l_1)\,\Gamma_{A^{f}\bar{c}^{h}c^{n}}(-l_1,l_1+p_1,-p_1)\,G_{c^{i}\bar{c}^{h}}(l_1+p_1,-l_1-p_1)
			\\
			&\hspace{1.2cm}\times\,\partial_t R_{\bar{c}^{b}c^{i}}(l_1+p_1,-l_1-p_1))
		\end{aligned}
		$}
\end{equation*}
\vspace{-0.5ex}

\noindent\hspace{3.0em}\includegraphics[width=0.6\linewidth]{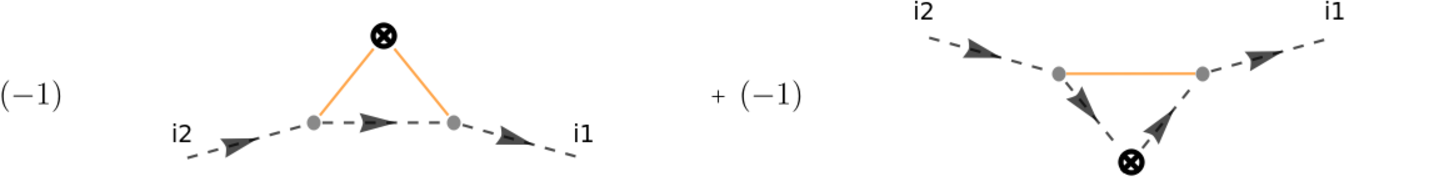}\\
To proceed, we require the explicit group structure of the ghost and gluon propagators, as well as the ghost-gluon vertex. We have already derived the necessary rules for the ghost vertices in \Cref{sec:Notation}. Assuming Landau gauge, a possible parametrisation of the gluon two-point vertex reads
\begin{align}
	\Gamma_{A_\mu^AA_\nu^B}(p,q) = 
	\left(
	Z_A(p)\Pi^\perp_{\mu\nu}(p) + \frac{1}{\xi}\Pi^\parallel_{\mu\nu}(p)
	\right)p^2\delta^{AB}
	\delta(p+q)\,.
\end{align}
Here, $\Pi^\perp_{\mu\nu}(p) = \delta_{\mu\nu} - \frac{p_\mu p_\nu}{p^2}$ and $\Pi^\parallel_{\mu\nu}(p) = \frac{p_\mu p_\nu}{p^2}$ are the transverse and longitudinal projectors, respectively.
In Landau gauge, the longitudinal part of the gluon propagator does not get renormalised and thus we have not added a dressing function to that part.

Altogether, we can write a set of Feynman rules as
\begin{mmaCell}{Input}
feynmanRulesYMGhost=\{
  \mmaDef{Propagator}[\{A,A\},\{\{p1_,\{v1_,c1_\}\},\{p2_,\{v2_,c2_\}\}\}]:>
    deltaAdjCol[c1,c2](1/ZA[Sqrt[sp[p2,p2]]]transProj[p2,v1,v2])/sp[p2,p2],
  \mmaDef{Propagator}[\{c,cb\},\{\{p1_,\{c1_\}\},\{p2_,\{c2_\}\}\}]:>
    -deltaAdjCol[c1,c2] 1/(Zc[Sqrt[sp[p2,p2]]]sp[p2,p2]),
  
  \mmaDef{GammaN}[\{A,cb,c\},\{\{p1_,\{v1_,c1_\}\},\{p2_,\{c2_\}\},\{p3_,\{c3_\}\}\}]:>-I vec[p2,v1]g FCol[c2,c3,c1],
  
  \mmaDef{Rdot}[\{A,A\},\{\{p1_,\{v1_,c1_\}\},\{p2_,\{v2_,c2_\}\}\}]:>
    deltaAdjCol[c1,c2]transProj[p2,v1,v2]RAdot[k^2,p^2],
  \mmaDef{Rdot}[\{cb,c\},\{\{p1_,\{c1_\}\},\{p2_,\{c2_\}\}\}]:>-deltaAdjCol[c1,c2]Rcdot[k^2,p^2]
 \};
\end{mmaCell}
We have already set $\xi=0$ for Landau gauge, so we don't include the longitudinal tensor structure in the gluon propagator.
Note that the default \FormTracer aliases for group and momentum tensors are used, if one does not specify them otherwise, e.g. \mathem{\mmaDef{sp}[p,q]} represents a scalar product $p\cdot q$, \mathem{\mmaDef{deltaAdjCol}[c2,c1]} stands in for a $\mathbbold{1}$ in adjoint colour space, etc. For a full list of all names \FormTracer knows, call \mathem{\mmaDef{ShowFormTracerDefinitions}[]}, which prints a full table thereof.

For the tracing of the flow, we have to project the tensor structures of the diagram. For this, we can use the projector onto the ghost two-point function calculated by \TensorBases. Then, we can also insert the rules into the expression and trace with the help of \FormTracer:
\begin{mmaCell}{Input}
\mmaDef{traceExprYMGhost} = \mmaDef{FTerm}[
  \mmaDef{TBGetProjector}["cbc",1,\{i1,i2\}/.\mmaDef{diagGhostFlow}["1-Loop"]["ExternalIndices"]],
  \mmaDef{diagGhostFlow}["1-Loop"]["Expression"]
 ]/.\mmaDef{feynmanRulesYMGhost}//\mmaDef{FormTrace}
\end{mmaCell}
\begin{mmaCell}{Output}
\{\mmaFrac{
-\mmaSup{g}{2} Nc RAdot[\mmaSup{k}{2},\mmaSup{p}{2}](\mmaSup{sp[l1,p1]}{2}-sp[l1,l1]sp[p1,p1])
}{
\mmaSup{sp[l1,l1]}{3}(sp[l1,l1]-2sp[l1,p1]+sp[p1,p1]) \mmaSup{ZA[\mmaSqrt{sp[l1, l1]}]}{2} Zc[\mmaSqrt{sp[l1,l1]-2sp[l1,p1]+sp[p1,p1]}]
},
\mmaFrac{
-\mmaSup{g}{2} Nc Rcdot[\mmaSup{k}{2},\mmaSup{p}{2}](\mmaSup{sp[l1, p1]}{2}-sp[l1,l1]sp[p1,p1])
}{
\mmaSup{sp[l1,l1]}\mmaSup{(sp[l1,l1]+2sp[l1,p1]+sp[p1,p1])}{2}ZA[\mmaSqrt{sp[l1,l1]}] \mmaSup{Zc[\mmaSqrt{sp[l1,l1]+2sp[l1,p1]+sp[p1,p1]}]}{2}
}\}
\end{mmaCell}
The result is the flow of the scalar part of $\Gamma_{\bar{c}c} = \frac{\delta}{\delta\bar{c}}\frac{\delta}{\delta c} \Gamma$.
Here, \mathem{\mmaDef{FormTrace}} delegates to the \FormTracer package, which sends the expression to \FORM, which efficiently performs all momentum and group algebra contractions. The result is either a scalar expression, or a list of scalar expressions for each diagram.

It is important to note that after tracing, momentum products like \mathem{\mmaDef{sp}[p,q]} will still be present in the output, and they need to be explicitly parametrised by the user before further processing.

%%%%%%%%%%%%%%%%%%%%%%%%%%%%%%
\subsection{Automatically creating diagrammatic rules}

To automatically generate a set of diagrammatic rules, \FunKit interfaces with the \TensorBases \cite{Braun:2025gvq} package, which provides the tensor expressions for interaction vertices. In \FunKit, a set of vertex bases is specified inside the \mathem{\mmaDef{setup}}.
\begin{mmaCell}{Input}
bases = <|\mmaDef{GammaN}->\{\{A,A\}->\{"AA",1\},
                   \{A,A,A\}->"AAAClass",
                   \{A,A,A,A\}->"AAAAClass",
                   \{A,cb,c\}->\{"Acbc",1\},
                   \{cb,c\}->"cbc"\},
          \mmaDef{S}->\{\{A,A\}->\{"AA",\{1\}\},
              \{A,A,A\}->"AAAClass", 
              \{A,A,A,A\}->"AAAAClass",
              \{A,cb,c\}->\{"Acbc",1\},
              \{cb,c\}->"cbc"\},
          \mmaDef{Propagator}->\{\{A,A\}->"AA",\{cb,c\}->"cbc"\},
          \mmaDef{Rdot}->\{\{A,A\}->\{"AA",1\},\{cb,c\}->"cbc"\}|>;
\mmaDef{setup} = <|"FieldSpace" -> \mmaDef{fields},
          "Truncation" -> \mmaDef{truncation},
          "FeynmanRules" -> bases|>;
\mmaDef{FSetGlobalSetup}[\mmaDef{setup}];
\end{mmaCell}
We have used in the above several default bases provided by \TensorBases. For more details on \TensorBases see \cite{Braun:2025gvq}. To see all available bases, one can call \mathem{\mmaDef{TBInfo[]}}.\footnote{In particular, to get acquainted with \TensorBases we point the reader to the example notebook at \url{https://github.com/satfra/TensorBases/blob/main/examples/Showcase.nb}}

Generally, there are three ways to specify vertex bases:
\begin{enumerate}[(i)]
	\item Giving only a name, e.g. \mathem{{A,A,A}->"AAAClass"}, which uses the full basis specified as \mathem{"AAAClass"}.
	\item Picking a subset of a larger basis, e.g. in QCD one could use \mathem{{A,qb,q}->{"AqbqDirect",1,4,7}} to pick out the most relevant basis elements 1, 4 and 7 from a standard quark-gluon tensor basis \mathem{"AqbqDirect"} as adapted from \cite{Cyrol:2017ewj, Braun:2025gvq}.
	\item Picking a subset of a larger basis and/or changing names of fields.
	As an example, for strange quarks \mathem{sb}, \mathem{s}, one could use \mathem{{A,sb,s}->{"AqbqDirectSF",1,4,7,s->q,sb->qb}} to change the names given to the fields in the pre-defined tensor basis \mathem{"AqbqDirectSF"}.
\end{enumerate}
With the above definition of \mathem{\mmaDef{setup}}, the diagrammatic rules can be automatically derived using
\begin{mmaCell}{Input}
diagRules = \mmaDef{FMakeDiagrammaticRules}[];
\end{mmaCell}
\begin{mmaCell}{Input}
\mmaDef{Propagator}[\{A,A\},\{\{p1,\{v1,c1\}\},\{p2,\{v2,c2\}\}\}]/.\mmaDef{FMakeDiagrammaticRules}[]
\end{mmaCell}
\begin{mmaCell}{Output}
\mmaFrac{deltaAdjCol[c2,c1] longProj[-p2,v2,v1]}{dressing[InverseProp,\{A,A\},2,\{p1,p2\}]} + \mmaFrac{deltaAdjCol[c2,c1] transProj[-p2,v2,v1]}{dressing[InverseProp,\{A,A\},1,\{p1,p2\}]}
\end{mmaCell}
The result of \mathem{\mmaDef{FMakeDiagrammaticRules}[]} is a replacement table that inserts all tensors in the specified basis for a given object, here the propagator. Each tensor is multiplied with a momentum-dependent \mathem{\mmaDef{dressing}} that can be later parametrised by the user. In the above output, for example, \FunKit uses the dressing of the inverse propagator in the diagrammatic rule for the propagator.

With this, one can directly obtain the explicit expressions for diagrams. For example, the 0-loop term of the gluon propagator DSE can be calculated as
\begin{mmaCell}{Input}
DSEAA = \mmaDef{FTakeDerivatives}[\mmaDef{FMakeDSE}[A[i1]], \{A[i2]\}]//\mmaDef{FTruncate}//\mmaDef{FRoute};
DSEAA["0-Loop"]["Expression"]
DSEAA["0-Loop"]["Expression"]/.\mmaDef{diagRules}
\end{mmaCell}
\begin{mmaCell}{Output}
FEx[FTerm[2,S[\{A,A\},\{\{-p1,\{v2,c2\}\},\{p1,\{v1,c1\}\}\}]]]
\end{mmaCell}
\begin{mmaCell}{Output}
FEx[FTerm[2,deltaAdjCol[c2,c1]dressing[S,\{A,A\},1,\{-p1,p1\}]transProj[p1,v2,v1]]]
\end{mmaCell}
Sometimes, a mixture of diagrammatic rules from given tensor bases, created with \mathem{\mmaDef{FMakeDiagrammaticRules}[]}, and hand-written additional rules may be appropriate. E.g. specifying the vertices of an O($N$) theory may be much easier when written out by hand, rather than creating separate tensor bases for every single vertex -- in particular, if $N=1$.

To obtain fully traced expressions, we usually need to use a projector to obtain the coefficient of a single element of the larger tensor basis. For the gluon propagator DSE, we can use the projector calculated by \TensorBases,
\begin{mmaCell}{Input}
projAA = \mmaDef{TBGetProjector}["AA",1,\{i1,i2\}/.\mmaDef{DSEAA}["1-Loop"]["ExternalIndices"]]
\end{mmaCell}
\begin{mmaCell}{Output}
\mmaFrac{deltaAdjCol[c1,c2] transProj[-p1,v1,v2]}{3 (-1 + \mmaSup{Nc}{2})}
\end{mmaCell}
Here, we made use of the \mathem{"ExternalIndices"} key to obtain a set of actual indices that \FunKit generated from the \mathem{"FieldSpace"} of the setup. With the projector, we can trace out the 1-loop diagrams of the gluon gap equation using \mathem{\mmaDef{FormTrace}}:
\begin{mmaCell}{Input}
DSEAAResult1L = \mmaDef{FormTrace}[
  \mmaDef{FTerm}[\mmaDef{projAA}, \mmaDef{DSEAA}["1-Loop"]["Expression"]]/.\mmaDef{diagRules}
 ];
\end{mmaCell}
Similarly, the above calculation can be also performed for all other diagrams appearing in the DSEs of the Yang--Mills system.

When using projectors for fields carrying Dirac indices, the superindices in the projector should be appropriately exchanged, so that a field contracts with its partner-field. For an illustration, see also the NJL example in \Cref{app:NJL}, which shows this at the example of fermions.

\begin{tcolorbox}[title=Algorithm details: Tracing,
	title filled=false,
	colback=pastelgreen!5!white,
	colframe=pastelgreen!73!black,
	fonttitle=\bfseries]
	
	To enable efficient tracing of large expressions using \FormTracer and \FORM, we automatically optimise the process in two directions:
	\begin{itemize}
		\item When giving multiple \mathem{\mmaDef{FTerm}} to \mathem{\mmaDef{FormTrace}} the terms are all traced separately in parallel.
		\item We make use of \FORM output optimisation. Although it is not available when outputting in the Mathematica format, we internally output the result as \textit{Fortran90} code, while using \FORM \texttt{O1} optimisation. This output is then parsed back to Mathematica code, and simplified in small chunks. The result is an efficient simplification of the very large expressions usually produced by \FORM.
	\end{itemize}
	
\end{tcolorbox}

%%%%%%%%%%%%%%%%%%%%%%%%%%%%%%
\subsection{Momentum projections}

\FunKit provides a few more tools to simplify certain tracing tasks. As an example, let us consider the DSE of the  ghost-gluon interaction, which we obtain as
\begin{mmaCell}{Input}
DSEAcbc = \mmaDef{FTakeDerivatives}[\mmaDef{FMakeDSE}[A[i1]], \{cb[i2],c[i3]\}]//\mmaDef{FTruncate}//\mmaDef{FPlot}//\mmaDef{FRoute};
\end{mmaCell}
\vspace{-1.0ex}

\noindent\hspace{3.0em}\includegraphics[width=0.85\linewidth]{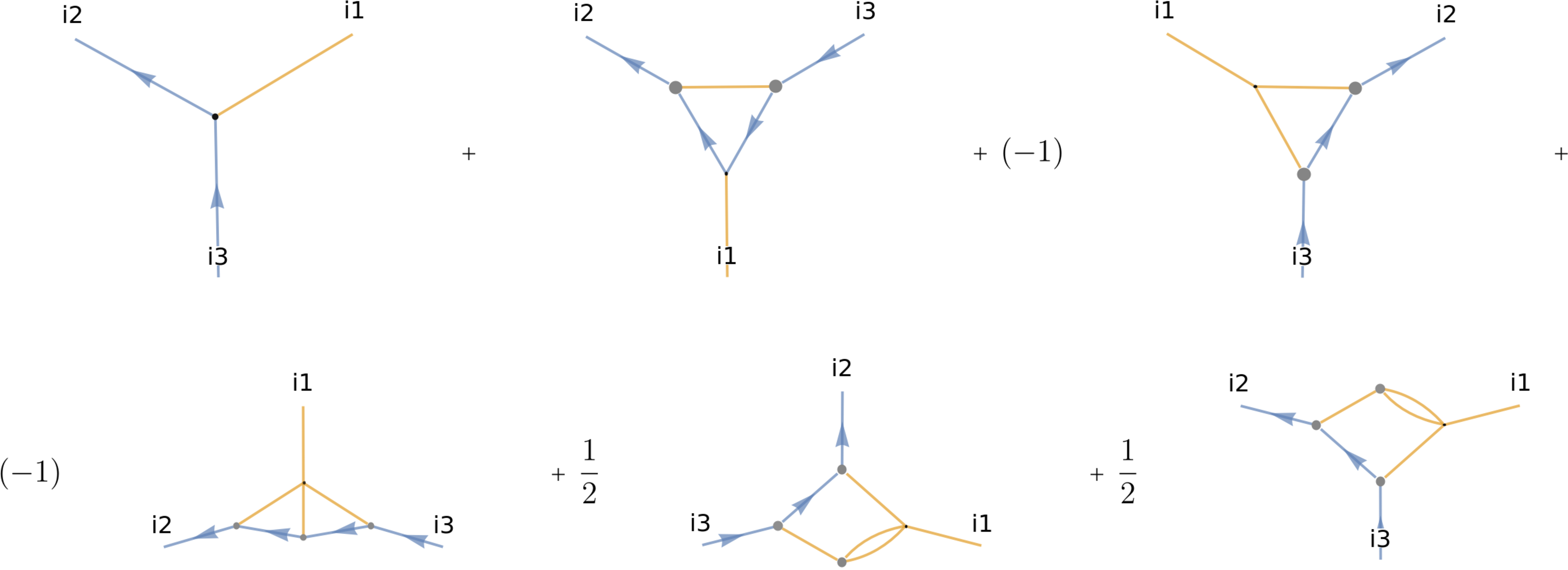}\\[1ex]
In the following, we trace the third diagram of the above.
In many situations, the symmetric-point approximation can be a good momentum configuration to obtain a reduced system of correlation functions. Specifically, if angular dependencies can be assumed to be mild, a system of symmetric-point diagrams defines a closed system of diagrams in DSE or fRG applications, see also \cite{Ihssen:2024miv, Eichmann:2014xya, Eichmann:2015nra, Eichmann:2025etg} for more explicit treatments of 3- to 5-point~functions.

The symmetric-point approximation effectively reduces the ($n-1$)-dimensional momentum dependence of each vertex dressing to a single scale. The approximation first consists of evaluating the flow of every diagram at a configuration specified through
\begin{align}
	p^2 = \frac{1}{n}\sum_{i=1}^n p_i^2,\qquad p_i\cdot p_j = - \frac{1}{n-1} p^2\,\,\textrm{for $i\neq j$}\,. 
	\label{eq:symmetricPoint}
\end{align}
An explicit choice for the $p_i$ is given by the centred ($n-1$)-simplex embedded in $d$ dimensions (or a lower-dimensional projection thereof, if $n>d$). 
In the case of the ghost-gluon vertex in four dimensions this would be vertices of a triangle
\begin{align}\notag 
	p_1 &= |p|\cdot \hat e (\pi/2, \pi/2, 0)\,,          	\\[1ex] \notag
	p_2 &= |p|\cdot \hat e (\pi/2, \pi/2, 2\pi/3)\,,    \\[1ex]
	p_3 &= |p|\cdot \hat e (\pi/2, \pi/2, 4\pi/3)\,, 
\end{align}
where $\hat e(\vartheta_1, \vartheta_2, \varphi)$ is the four-dimensional unit vector in hyperspherical coordinates. Following that, wherever the ghost-gluon dressing $\lambda_{c\bar{c}A}$ appears in a diagram, it is evaluated at the average momentum
\begin{align}
	\lambda_{c\bar{c}A}(\bar{p}) = \lambda_{c\bar{c}A}\left(\sqrt{\frac{p_1^2+p_2^2+p_3^2}{3}}\right)\,.
\end{align}
To trace the diagram, we once again use a projector taken from the \TensorBases library. Using the \mathem{\mmaDef{FMakeSPFormRule}} function, we create a \FORM rule that can be passed to \mathem{\mmaDef{FormTrace}} as a post-processing rule:
\begin{mmaCell}{Input}
projAcbc = \mmaDef{TBGetProjector}["Acbc",1,\{i1,i2,i3\}/.\mmaDef{DSEAcbc}["1-Loop"]["ExternalIndices"]];
formRuleSP = \mmaDef{FMakeSPFormRule}[\{l1\},p,\{p1,p2,p3\}];
\mmaDef{FormTrace}[\mmaDef{FTerm}[projAcbc]**\mmaDef{DSEAcbc}["1-Loop"]["Expression"][[2]]/.\mmaDef{diagRules},\{\},formRuleSP]
\end{mmaCell}
\begin{mmaCell}{Output}
(Nc (-1+2 cos[p1,l1]cos[p2,l1]+2 cos[p2,l1]^2)dressing[GammaN,\{A,cb,c\},1,\{-l1,l1+p1+p2,-p1-p2\}]
 dressing[GammaN,\{A,cb,c\},1,\{l1,p2,-l1-p2\}] dressing[S,\{A,cb,c\},1,\{p1,l1+p2,-l1-p1-p2\}]
  (2 cos[p1,l1] Sqrt[sp[l1,l1]]+4 cos[p2,l1] Sqrt[sp[l1,l1]]+3 Sqrt[sp[p,p]]) Sqrt[sp[p,p]]) /
 (12 dressing[InverseProp,\{A,A\},1,\{l1,-l1\}] dressing[InverseProp,\{cb,c\},1,\{l1+p2,-l1-p2\}]
  dressing[InverseProp,\{cb,c\},1,\{l1+p1+p2,-l1-p1-p2\}])
\end{mmaCell}
%
%The explicit \FORM code generated by the call to \mathem{\mmaDef{FMakeSPFormRule}} reads
%%
%\begin{mmaCell}{Input}
%\mmaDef{formRuleSP}
%\end{mmaCell}
%\begin{mmaCell}{Output}
%\{"Vector p1,p2,p3, l1 ,p;
%Symbol n;
%AutoDeclare CFunction cos;
%Set exMom : p1,p2,p3;
%Set loopMom : l1;
%	
%#define SPOrd "3"
%	
%*** project to the SP
%#procedure ProjSP()
%id FTxsp(p1?exMom,p1?exMom)^-1 = FTxsp(p,p)^-1;
%id FTxsp(p1?exMom,p1?exMom) = FTxsp(p,p);
%id FTxsp(p1?exMom,p2?exMom)^-1 = (-FTxsp(p,p)/(`SPOrd'-1))^-1;
%id FTxsp(p1?exMom,p2?exMom) = -FTxsp(p,p)/(`SPOrd'-1);
%	
%id FTxsp(p1?exMom,l1?loopMom)^-1 = (sqrt(FTxsp(p,p))*sqrt(FTxsp(l1,l1))*cos(p1,l1))^-1;
%id FTxsp(p1?exMom,l1?loopMom) = (sqrt(FTxsp(p,p))*sqrt(FTxsp(l1,l1))*cos(p1,l1));
%#endprocedure
%	
%#call SCALL(ProjSP)
%.sort"\}
%\end{mmaCell}
%
Note that \mathem{\mmaDef{FMakeSPFormRule}} simply generates \FORM code that gets injected at the end of the tracing pipeline.
The leftover cosines \mathem{\mmaDef{cos}[p1,l1]}, as well as the other momentum arguments have to be parametrised by the user before actual evaluation or code export.

A full example, deriving a set of DSEs and flow equations for Yang--Mills theory with the above specified truncation, can be found in \texttt{examples/Yang-Mills.nb} in the \FunKit repository.

%%%%%%%%%%%%%%%%%%%%%%%%%%%%%%
\subsection{Defining new bases}
\label{sec:defbases}

Beyond the pre-defined bases shipped with \TensorBases, users can define custom tensor bases via the \mathem{\mmaDef{TBConstructBasis}} function of the \TensorBases package \cite{Braun:2025gvq}. A complete worked example, deriving the fRG flow equations for an NJL model with user-defined Dirac bases for two- and four-fermion interactions, is given in \Cref{app:NJL}. The example demonstrates basis construction, setup definition, diagrammatic rule generation, and the full tracing pipeline.

%%%%%%%%%%%%%%%%%%%%%%%%%%%%%%
\section{Automatic Code Generation}
\label{sec:COEN}

\FunKit provides a set of tools to conveniently translate the Mathematica expressions resulting after performing all traces into code that can be directly incorporated into a numerical program.
To that end, \FunKit provides direct output to Julia and Fortran functions, as well as a set of functions to generate C++ classes and methods with arbitrary types by means of templates.

Before exporting the output of tracing, i.e., after the steps detailed in \Cref{sec:TRACY}, the user must parametrise all remaining symbolic objects (dressings, scalar products, angular variables). This is necessary to obtain numerically tractable expressions. For the Yang--Mills system we have taken also previously as an example, such a parametrisation can be given by
\begin{mmaCell}{Input}
parametrizations=\{
	\mmaDef{dressing}[InverseProp,\{A,A\},1,\{p1_,p2_\}]:>ZA[Sqrt[sp[p2,p2]]]sp[p2,p2],
	\mmaDef{dressing}[InverseProp,\{A,A\},2,\{a_,b_\}]:>xi^-1,
	\mmaDef{dressing}[InverseProp,\{cb,c\},1,\{p1_,p2_\}]:>-Zc[Sqrt[sp[p2,p2]]]sp[p2,p2],
	\mmaDef{dressing}[S,\{A,A,A\},1,\{p1_,p2_,p3_\}]:>gS,
	\mmaDef{dressing}[S,\{A,A,A,A\},1,\{p1_,p2_,p3_,p4_\}]:>gS^2,
	\mmaDef{dressing}[S,\{A,cb,c\},1,\{p1_,p2_,p3_\}]:>gS,
	\mmaDef{dressing}[GammaN,\{A,A,A\},1,\{p1_,p2_,p3_\}]:>ZA3[Sqrt[(sp[p1,p1]+sp[p2,p2]+sp[p3,p3])/3]],
	\mmaDef{dressing}[GammaN,\{A,cb,c\},1,\{p1_,p2_,p3_\}]:>ZAcbc[Sqrt[(sp[p1,p1]+sp[p2,p2]+sp[p3,p3])/3]],
	lf1->l1,
	sp[l1,l1]:>l1^2, sp[p1,p1]:>p^2, sp[l1,p1]:>cospl1 l1 p,
	cos[p,l1]->cospl1,
 \};
parametrize[expr_]:=\mmaDef{UseLorentzLinearity}[expr//.parametrizations]//.parametrizations//\mmaDef{Simplify};
\end{mmaCell}
In the above, we have parametrised all \mathem{\mmaDef{dressing}} functions and the remaining momentum products. With this in hand, we can immediately obtain Julia code from our previous tracing of the DSE of the gluon propagator:
\begin{mmaCell}{Input}
\mmaDef{MakeJuliaFunction}[
  \mmaDef{Total}@\mmaDef{DSEAAResult1L}//\mmaDef{parametrize},
  "Parameters" -> \{"l1", "cospl1", "p", "Nc", "gS", "ZA", "Zc", "ZA3", "ZAcbc", "ZA4"\},
  "Name" -> "DSE_kernel_AA_1Loop"
 ]
\end{mmaCell}
\begin{mmaCell}{Output}
function DSE_kernel_AA_1Loop(l1, cospl1, p, Nc, gS, ZA, Zc, ZA3, ZAcbc, ZA4)
  _interp1 = ZA(sqrt(l1^2))
  _interp2 = ZA(sqrt(l1^2 + 2*cospl1*l1*p + p^2))
  _interp3 = ZA3(sqrt(0.6666666666666666)*sqrt(l1^2 + cospl1*l1*p + p^2))
  _interp4 = ZAcbc(sqrt(0.6666666666666666)*sqrt(l1^2 + cospl1*l1*p + p^2))
  _interp5 = Zc(sqrt(l1^2))
  _interp6 = Zc(sqrt(l1^2 + 2*cospl1*l1*p + p^2))
  _cse1 = cospl1^2
  _cse2 = -1 + _cse1
  _cse4 = l1^2
  _cse3 = 1/_cse4
  _cse5 = 2*cospl1*l1*p
  _cse6 = p^2
  _cse7 = _cse4 + _cse5 + _cse6
  _cse8 = 1/_interp1
  return fma(6,(_cse2*_cse3*_cse4^2*_cse8*_interp3*gS*Nc)/(_cse7^2*_interp2),fma(16,(_cse2*_cse3*_cse4*_cse6*_cse8*_interp3*gS*Nc)/(_cse7^2*_interp2),fma(2,(_cse1*_cse2*_cse3*_cse4*_cse6*
    _cse8*_interp3*gS*Nc)/(_cse7^2*_interp2),fma(6,(_cse2*_cse3*_cse6^2*_cse8*_interp3*gS*Nc)/(_cse7^2*_interp2),fma(-1,(_cse2*_interp4*gS*Nc)/(_cse7*_interp5*_interp6),fma(7,_cse3*_cse8*
    gS^2*Nc,fma(-1,_cse1*_cse3*_cse8*gS^2*Nc,fma(12,(_cse2*_cse3*_cse8*_interp3*cospl1*gS*l1^3*Nc*p)/(_cse7^2*_interp2),fma(12,(_cse2*_cse3*_cse8*_interp3*cospl1*gS*l1*Nc*p^3)/(_cse7^2*
    _interp2),0)))))))))/3.
end
\end{mmaCell}
This code can then be copied into an existing Julia code and called by integration routines and iterators to solve the system of DSEs.

Similarly, C++ code can be generated using the function \mathem{\mmaDef{MakeCppFunction}}. However, this function permits more detailed configuration of parameter and return types. Without specifying any of these, \cpp{auto} is simply used and deducing types is left to the compiler. Note that the code thus generated requires at least the C++20 standard.
Furthermore, we add \cpp{using namespace std;} manually at the start of the function body. This makes sure that all mathematical operations are defined (additionally the \cpp{"cmath"} header should be included). However, a user may define these functions differently, e.g. using \cpp{cuda::std} or manually.
\begin{mmaCell}{Input}
\mmaDef{MakeCppFunction}[
  \mmaDef{Total}@\mmaDef{DSEAAResult1L}//\mmaDef{parametrize},
    "Parameters" -> \{"l1", "cospl1", "p", "Nc", "gS", "ZA", "Zc", "ZA3", "ZAcbc", "ZA4"\},
    "Name" -> "DSE_kernel_AA_1Loop",
    "Body" -> "using namespace std;"
 ]
\end{mmaCell}
\begin{mmaCell}{Output}
auto DSE_kernel_AA_1Loop(const auto &l1, const auto &cospl1, const auto &p, const auto &Nc,
                         const auto &gS, const auto &ZA, const auto &Zc, const auto &ZA3, 
                         const auto &ZAcbc, const auto &ZA4)
\{
  using namespace std;
  const auto \_interp1 = ZA(sqrt(powr<2>(l1)));
  const auto \_interp2 = ZA(sqrt(powr<2>(l1) + 2. * cospl1 * l1 * p + powr<2>(p)));
  const auto \_interp3 = ZA3(0.816496580927726 * sqrt(powr<2>(l1) + cospl1*l1*p + powr<2>(p)));
  const auto \_interp4 = ZAcbc(0.816496580927726 * sqrt(powr<2>(l1) + cospl1*l1*p + powr<2>(p)));
  const auto \_interp5 = Zc(sqrt(powr<2>(l1)));
  const auto \_interp6 = Zc(sqrt(powr<2>(l1) + 2. * cospl1 * l1 * p + powr<2>(p)));
  const auto \_cse1 = powr<2>(cospl1);
  const auto \_cse2 = -1. + \_cse1;
  const auto \_cse4 = powr<2>(l1);
  const auto \_cse3 = powr<-1>(\_cse4);
  const auto \_cse5 = 2. * cospl1 * l1 * p;
  const auto \_cse6 = powr<2>(p);
  const auto \_cse7 = \_cse4 + \_cse5 + \_cse6;
  const auto \_cse8 = powr<-1>(\_interp1); // clang-format off

  return 0.3333333333333333 * fma(6.,
      \_cse2 * \_cse3 * powr<2>(\_cse4) * powr<-2>(_cse7) * _cse8 * powr<-1>(_interp2) * _interp3 *
        gS * Nc, fma(16.,
          _cse2 * _cse3 * _cse4 * _cse6 * powr<-2>(_cse7) * _cse8 * powr<-1>(_interp2) *
          _interp3 * gS * Nc, fma(2.,
            _cse1 * _cse2 * _cse3 * _cse4 * _cse6 * powr<-2>(_cse7) * _cse8 *
              powr<-1>(_interp2) * _interp3 * gS * Nc, fma(6.,
                _cse2 * _cse3 * powr<2>(_cse6) * powr<-2>(_cse7) * _cse8 * powr<-1>(_interp2) *
                _interp3 * gS * Nc, fma(-1.,
                  _cse2 *powr<-1>(_cse7) * _interp4 * powr<-1>(_interp5) *
                  powr<-1>(_interp6) * gS * Nc, fma(7., _cse3 * _cse8 * powr<2>(gS) * Nc,
                    fma(-1., _cse1 * _cse3 * _cse8 * powr<2>(gS) * Nc, fma(12.,
                      _cse2 * _cse3 * powr<-2>(_cse7) * _cse8 * powr<-1>(_interp2) *
                        _interp3 * cospl1 * gS * powr<3>(l1) * Nc * p, fma(12.,
                          _cse2 * _cse3 * powr<-2>(_cse7) * _cse8 *
                          powr<-1>(_interp2) * _interp3 * cospl1 * gS * l1 * Nc *
                          powr<3>(p), 0.)))))))));
  // clang-format on
\}
\end{mmaCell}
Fortran code can be generated analogously using the function \mathem{\mmaDef{MakeFortranFunction}}, which produces modern free-form Fortran with \texttt{implicit none} and typed parameter declarations. The expression serialiser \mathem{\mmaDef{FortranCodeForm}} translates Mathematica expressions to modern free-form Fortran and generates \texttt{double precision} functions with \texttt{implicit none}, making it suitable for integration into existing Fortran-based numerical codebases.

By default, the code generation targets a register budget of 32 registers per kernel, reflecting typical GPU hardware constraints. For CPU-targeted applications, where register pressure is less restrictive, the user may adjust this setting via the \mathem{\mmaDef{FSetRegisterSize}} function. For C++, note also that \FunKit uses a custom function to describe integer powers, see \Cref{app:powr}.

The code generation of \FunKit targets C++, Julia and Fortran, as these languages offer the performance characteristics required for realistic numerical applications.
In particular, the generated code is designed for tight integration with high-performance solvers and GPU kernels, where languages such as Python would introduce prohibitive overhead.
Should the need arise, the modular design of \FunKit makes the addition of further backends straightforward.

\begin{tcolorbox}[title=Algorithm details: Code generation,
	title filled=false,
	colback=pastelgreen!5!white,
	colframe=pastelgreen!73!black,
	fonttitle=\bfseries]
	
	The Julia, C++ and Fortran outputs of \FunKit process expressions through the following pipeline:
	\begin{enumerate}
		\item \textbf{Interpolator hoisting (global):} Extract all calls matching patterns that could be interpolator or function calls and assign them to temporaries, eliminating redundant global-memory reads.
		\item \textbf{Split decision (global):} If the total weighted complexity exceeds a threshold, partition the expression into sub-kernels; shared interpolators are kept in a common scope.
		\item \textbf{Common subexpression elimination (per kernel):} We identify common sub-expressions and assign each to a \cpp{_cseN} temporary.
		\item \textbf{Power normalisation (per kernel):} Rewrite integer powers of expressions in terms of already-hoisted temporaries, e.g.\ \cpp{powr<-4>(l1)} becomes \cpp{powr<-2>(_cse1)} when \cpp{_cse1 = powr<2>(l1)} exists.
		\item \textbf{Algebraic factoring (per kernel):} Apply \mathem{FactorTerms} (up to three passes) to group additive terms sharing common factors.
		\item \textbf{Transcendental hoisting (per kernel):} Hoist expensive calls (e.g. \cpp{Exp}, \cpp{Log}, \cpp{Sin}, \cpp{Cos}, \cpp{Sqrt}, \ldots) to temporaries; if both \cpp{Exp[x]} and \cpp{Exp[-x]} appear, compute one and derive the other via reciprocal.
		\item \textbf{FMA restructuring (per kernel):} Detect \cpp{a*b + c} patterns and rewrite them as fused multiply-add groups, enabling single-instruction FMA execution.
		\item \textbf{Code formatting:} Emit temporaries as \cpp{const auto _cseN = <expr>;} and the final value as a \cpp{return} statement. Sub-kernels are wrapped in separate scopes that accumulate into a shared result~variable.
	\end{enumerate}
	The total number of available slots for hoisting of common sub-expressions per kernel can be set using \mathem{\mmaDef{FSetRegisterSize[int]}}, with a default of 32.
	
\end{tcolorbox}

%%%%%%%%%%%%%%%%%%%%
\section{Implementation example: Yang--Mills theory}
\label{sec:YMExample}

Using the \DiFfRG \cite{Sattler:2024ozv} numerical framework, we provide an implementation example for a Yang--Mills theory with full momentum dependences in the \FunKit repository \cite{FunKitRepo} at \texttt{examples/Yang-Mills/}.

The corresponding notebook at \texttt{examples/Yang-Mills/Yang-Mills.nb} uses \FunKit and \TensorBases to derive the flow equations. 
The flow equations are built into a C++ library, which is assembled from the Mathematica expressions using the code export facilities of \FunKit that are called and assembled by an interface layer provided by \DiFfRG.
These flow equations are then assembled into the full integro-differential system in a model specification \texttt{examples/Yang-Mills/model.hh} which can be solved by \DiFfRG. To build this example, the user should install first \DiFfRG, e.g. by the provided install script, then create a folder inside \texttt{examples/Yang-Mills/} and build:
\begin{lstlisting}[language=Bash]
mkdir build
cd build
cmake ..
make
\end{lstlisting}
The simulation can be then run by
\begin{lstlisting}[language=Bash]
./YangMills
\end{lstlisting}
All results are exported to a HDF5 file \texttt{output.hdf5}, which contains the calculated correlation functions on the corresponding momentum grids.

For the derivation, we project the full vertices of all classical tensor structures in Landau gauge onto the symmetric point, as explained in \Cref{sec:TRACY}. 
The initial condition has been adjusted such that the STIs are fulfilled in the ultraviolet, and the scaling solution is obtained in the infrared. \DiFfRG then integrates the flow equations over six orders of magnitude.
The details of the truncation and a more in-depth explanation of its features can be found in \cite{Cyrol:2016tym}. With this computation we essentially reproduce \cite{Cyrol:2016tym}.

\begin{figure*}[t]
	\centering
	\hspace{-0.9cm}\includegraphics[width=1.05\linewidth]{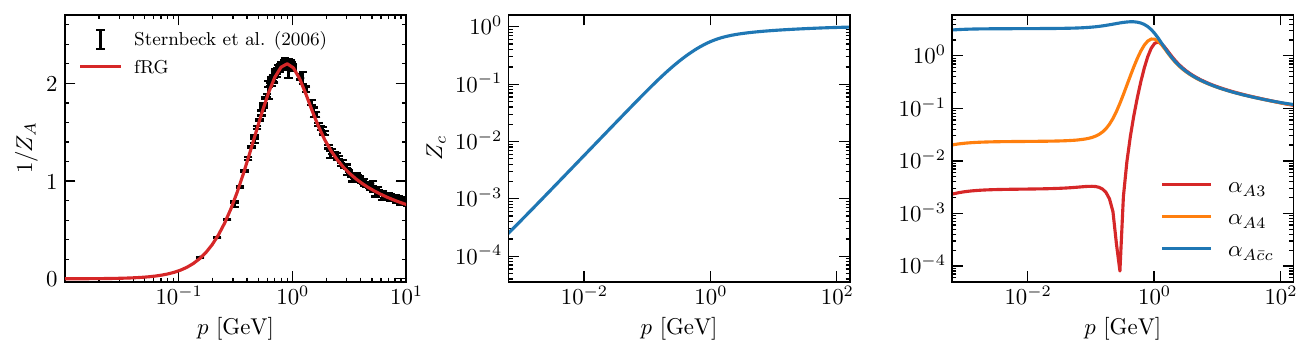}
	
	\caption{Results of the Yang--Mills example in \texttt{examples/Yang-Mills/}.
		Left: Inverse gluon dressing $1/Z_A(p)$ compared to lattice data from \cite{Sternbeck:2006cg}.
		Center: Ghost dressing in the scaling limit.
		Right: Avatars of the strong coupling.
	}
	\label{fig:YM}
\end{figure*}

We show the numerical results of this calculation in \Cref{fig:YM}. There, the gluon propagator is shown in comparison to the lattice data from \cite{Sternbeck:2006cg}, and we also show the ghost dressing which exhibits scaling down to $10^{-3}\,\textrm{GeV}$. 
Finally, we show the avatars of the strong coupling, which agree down to about $2\,\textrm{GeV}$, where they start to differ substantially in the infrared.

%%%%%%%%%%%%%%%%%%%%
\section{Comparisons and Benchmarks}
\label{sec:comparisons}

In the following, we compare \FunKit (v1.0.0) with the two most widely used packages for deriving functional equations, \DoFun (v3.0)~\cite{Huber:2019dkb,Huber:2011qr} and \QMeS (v1.2)~\cite{Pawlowski:2021tkk}.                                         

\Cref{tab:comparison} gives a feature-by-feature overview. All three packages support the derivation of DSEs and fRG flow equations for theories with any kind of field content. 
Beyond this common ground, \FunKit additionally supports arbitrary user-defined master equations, multi-index functional derivatives, source fields and multi-loop momentum routing.
Conversely, \DoFun additionally provides the derivation of composite operator equations -- however, these can be readily implemented in \FunKit using expressions with explicit \mathem{\mmaDef{FDOp}}, showing the strength of the vocabulary used by \FunKit. This is explicitly shown in the example \texttt{examples/CompositeOperators.nb}.

\FunKit also provides utilities not present in either \DoFun or \QMeS, e.g. integration of \FORM-based tracing with caching through \FormTracer, specification of tensorial bases for diagrammatic rules through \TensorBases, \LaTeX{} equation rendering and optimised C++, Julia and Fortran code generation. 
An earlier package that supported C++ code generation is \texttt{CrasyDSE} \cite{Huber:2011xc}, which was however specifically developed for DSEs.

Another special feature of \FunKit is the ability to identify topologically identical diagrams at any loop order, including disconnected ones. To do so, \FunKit traverses graphs in any possible sequence of paths to establish topological identity and relative signs, see also the green box above for Simplification.

Furthermore, for interoperability \FunKit includes explicit format conversion layers for both \DoFun and \QMeS output, allowing straightforward integration into existing pipelines and simplifying first contact for users.

\begin{figure*}
	\centering
	\begin{minipage}{0.92\textwidth}
		\includegraphics[width=1.05\textwidth]{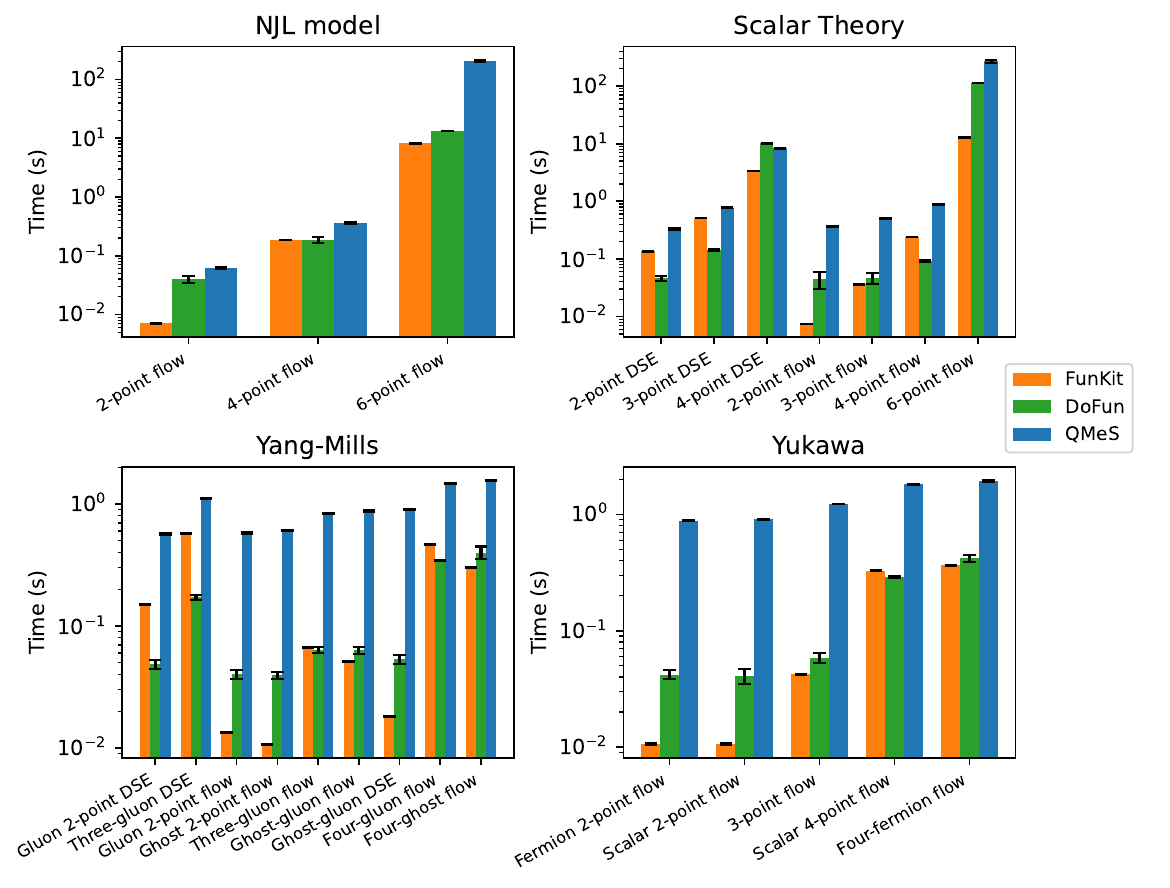}
	\end{minipage}
	\caption{Performance comparison between \FunKit (v1.0.0), 
		\DoFun (v3.0)~\cite{Huber:2019dkb,Huber:2011qr} and \QMeS (v1.2)~\cite{Pawlowski:2021tkk}.
		In all cases, we measured the total time to take the derivatives and truncate the expression.
		Especially for very large expressions with several thousand diagrams, e.g. the six-point and four-point flows, \FunKit excels.
		All benchmarks were performed on a 16-core AMD Ryzen 9 7945HX machine.
		Two warmup-cycles were used before five benchmark runs, which led to the error bars as shown in the plots. }
	\label{fig:benchmarks}
\end{figure*}

In \Cref{fig:benchmarks}, we show timing benchmarks for the derivation and truncation of correlation functions in four theories of increasing complexity: 
a four-fermion (NJL-type) model, a scalar $O(N)$ model, a Yukawa theory, and a Yang--Mills theory. In each case, the total wall time for taking the functional derivatives and truncating the resulting expression is measured. These benchmarks were performed on a 16-core AMD Ryzen 9 7945HX machine. Parallelisation was handled internally by each library.
The benchmark scripts can be found in the \FunKit repository in the \texttt{benchmarks/} folder, together with a visualisation Python script.

For small expressions, such as the scalar two-point function, all three packages perform well below a tenth of a second; however, \FunKit even proves to be an order of magnitude faster.
As the number of fields and the size of the truncation grow, however, \FunKit consistently outperforms \QMeS, while showing parity with \DoFun.
With even larger expressions, \FunKit is also able to outperform \DoFun: this is particularly pronounced for expressions involving many diagrams, such as the six-fermion flow equation in the four-fermion model, or the six-scalar flow in the scalar theory, where \FunKit can be faster by up to an order of magnitude.

It is important to note that even though \QMeS is slower than \DoFun or \FunKit, its most important advantage is its relative simplicity. This is achieved through straightforwardly implementing the superindex notation used also by \FunKit without any further optimisations that could introduce errors, with the aim of reliable handling of (fermionic) signs.

\begin{table*}[t]
	\centering
	\renewcommand{\arraystretch}{1.25}
	\renewcommand*\footnoterule{}
	\begin{minipage}{1\linewidth}
		\small
		\hspace{-0.5cm}
		\begin{tabular}[t]{l c c c |}
			\toprule
			\textbf{Feature} & \FunKit & \DoFun & \QMeS \\
			\midrule
			\multicolumn{3}{l}{\textit{Master equations}} &\\
			\quad Dyson--Schwinger equations             & \cmark  & \cmark  & \cmark \\
			\quad Flow equations (fRG)                  & \cmark  & \cmark  & \cmark \\
			\quad (Modified) Slavnov--Taylor id.         & \cmark  & \xmark  & \cmark \\
			\quad Composite operator equations          & (\cmark)\footnote{Although \FunKit does not provide these natively, they can be easily implemented by hand by constructing the objects with \texttt{FDOp}, see the provided example in \Cref{tab:examples}.}  & \cmark  & \xmark \\
			\quad Arbitrary master equations            & \cmark  & \xmark  & \xmark \\
			\quad Multi-index func.\ derivatives        & \cmark  & \xmark  & \xmark \\
			\hline
			\multicolumn{3}{l}{\textit{Field content}} &\\
			\quad Commuting (bosonic) fields            & \cmark  & \cmark  & \cmark \\
			\quad Grassmann (fermionic) fields          & \cmark  & \cmark  & \cmark \\
			\quad Complex fields                        & \cmark  & \cmark  & \cmark \\
			\quad Background / source fields            & \cmark  & \xmark  & (\cmark)\footnote{\QMeS supports only BRST sources.} \\
			\hline
			\multicolumn{3}{l}{\textit{Truncation \& simplification}} & \\
			\quad Vertex truncation                     & \cmark  & \cmark  & \cmark \\
			\quad Loop-order filtering                  & \cmark  & \cmark  & \xmark \\
			\quad Broken-symmetry phase                 & \cmark  & \cmark  & \xmark \\
			\quad Diagram topology identification       & \cmark  & (\cmark)\footnote{\DoFun can identify diagrams up to two-loop.}  & (\cmark)\footnote{\QMeS can identify diagrams up to one-loop.} \\
%			\bottomrule
		\end{tabular}%
		\begin{tabular}[t]{l c c c}
			\toprule
			\textbf{Feature} & \FunKit & \DoFun & \QMeS \\
			\midrule
			\multicolumn{4}{l}{\textit{Momentum routing \& diagrammatic rules}} \\
			\quad One-loop momentum routing             & \cmark  & \cmark  & \cmark \\
			\quad Multi-loop momentum routing           & \cmark  & (\cmark)\footnote{\DoFun can route diagrams up to two-loop.}  & \xmark \\
			\quad Feynman rule generation               & \cmark  & \cmark  & \xmark \\
			\quad Tensor basis specification            & \cmark  & \xmark  & \xmark \\
			\midrule
			\multicolumn{4}{l}{\textit{Index \& trace contraction}} \\
			\quad Tracing of group structures           & \cmark  & \xmark  & \xmark \\
			\midrule
			\multicolumn{4}{l}{\textit{Output \& visualisation}} \\
			\quad Diagram plots                         & \cmark  & \cmark  & \xmark \\
			\quad \LaTeX{} equation output              & \cmark  & \xmark  & \xmark \\
			\midrule
			\multicolumn{4}{l}{\textit{Code generation}} \\
			\quad C\texttt{++} / Julia / Fortran                 & \cmark  & \xmark  & \xmark \\
			\midrule
			\multicolumn{4}{l}{\textit{Interoperability}} \\
			\quad \texttt{FeynCalc} compatibility                & \xmark  & (\cmark)\footnote{\DoFun provides a utility function to safely load the \texttt{FeynCalc} package.} & \xmark \\
			\quad Cross-package format conversion       & \cmark  & \xmark  & \xmark \\
%			\bottomrule
		\end{tabular}
	\end{minipage}
	
	\caption{Feature comparison of \FunKit (v1.0.0), \DoFun (v3.0)~\cite{Huber:2019dkb,Huber:2011qr} and \QMeS (v1.2)~\cite{Pawlowski:2021tkk}. A \cmark{} indicates full support, (\cmark) partial or indirect support, and \xmark{} no support.
		Note that \FunKit has, a priori, a broader scope than the other packages, so features not present for \DoFun or \QMeS do not necessarily represent deficiencies.}
	\label{tab:comparison}
\end{table*}

%%%%%%%%%%%%%%%%%%%%
\section{Conclusions}

In this work, we have presented \FunKit, a Mathematica package for the derivation and processing of functional equations in quantum field theory.
It provides tools for tasks from the generation of algebraic expressions to the output of code that can be directly used in numerical applications.
In particular, we focus on a flexible and easy-to-use framework for deriving Dyson--Schwinger equations and functional Renormalisation Group equations for general quantum field theories. The package is designed to be extensible, allowing users to define their own field content, interactions, and truncations, and its modular structure readily accommodates future changes and improvements.

The correctness of the \FunKit package is verified by an extensive test suite, comparing with other popular packages such as \DoFun, \QMeS and analytical expressions from the literature.

In comparison to \DoFun or \QMeS, \FunKit allows both for broader applications, not being restricted to particular master equations. As an example for such an application, we have also provided an example notebook deriving flows with generalised flow equations, see \Cref{tab:examples}.
It also has a broader scope, as it culminates in either fully traced analytic expressions or code ready for use in a numerical framework. To our knowledge, no other package provides the entire pipeline depicted in \Cref{fig:workflow}.

We have also described the algorithms used in \FunKit, which are optimised at every stage: parallelisation is dynamically switched on and off throughout the workflow to maximise performance, and the generated code is highly optimised for both CPU and GPU applications.

We demonstrated the capabilities of \FunKit using the example of Yang--Mills theory, scalar field theory, a Yukawa theory and an NJL model. As for master equations, we show DSEs, the Wetterich equation, the modified STIs and dynamically reparametrised flow equations. All these examples can be found in the \FunKit repository, see also \Cref{tab:examples}.

\FunKit covers a broad range of applications, but of course some limitations remain.
A fundamental challenge is the resolution of functional derivatives, which involves a product-rule expansion that can lead to rapid intermediate expression growth for large truncations.
Although \FunKit mitigates this through in-loop simplification, parallelisation and appropriate algorithm design, the scaling remains a practical bottleneck for very large systems. This is however not a problem specifically of \FunKit, but rather of the approach itself.

Furthermore, \FunKit assumes translation invariance: momentum conservation is imposed at every vertex and each field carries a single momentum. Settings that break this -- e.g. Bloch momenta, finite-volume backgrounds or Wigner-transformed fields -- have to encode the additional structure via auxiliary indices.
More exotic field content, e.g. parastatistics, is not supported either, but could be accommodated by generalising the field-space metric.

Also, while \FunKit generates optimised code for the integrands of functional equations, it does not provide a numerical integration framework; the user must embed the generated code into their own solver or use an existing numerical framework such as \DiFfRG \cite{Sattler:2024ozv}.

\FunKit is already a basis for several on-going projects in the fQCD collaboration \cite{fQCD} involving large QCD truncations with dynamical hadronisation and other flowing reparametrisations.
As \FunKit is embedded in the tools of the fQCD collaboration, it has also become the symbolic backend of the \DiFfRG framework \cite{Sattler:2024ozv}, and will continue to see improvements in the future.

%%%%%%%%%%%%%%%%%%%%
\section*{Acknowledgments}
We thank Maurice Brezavsek, Andreas Gei\ss{}el, Friederike Ihssen, Keiwan Jamaly, Konrad Kockler, Ugo Mire, Jan~M. Pawlowski and Jonas Wessely for discussions, finding bugs, proofreading this work, and collaboration on related topics. This work is done within the fQCD collaboration \cite{fQCD} and we thank its members for discussions and collaborations on related projects.
FRS acknowledges funding by the GSI Helmholtzzentrum f\"ur Schwerionenforschung and by HGS-HIRe for FAIR. 
FRS is supported by the Deutsche Forschungsgemeinschaft (DFG, German Research Foundation) through the Emmy Noether Programme Project No. 54526179.

\vspace{1cm}

%%%%%%%%%%%%%%%%%%%%

%\newpage
\appendix

\gdef\thesection{\Alph{section}}
\makeatletter
\renewcommand\@seccntformat[1]{\appendixname\ \csname the#1\endcsname.\hspace{0.5em}}
\makeatother

\renewcommand{\thefigure}{\Alph{section}.\arabic{figure}}
\setcounter{figure}{0}

\begingroup
\allowdisplaybreaks

%%%%%%%%%%%%%%%%%%%%%%%%%%%%%%
\section{Superindex Notation Details}
\label{app:notation_details}

In this appendix, we provide some additional details and derivations for the superindex notation summarised in \Cref{sec:Notation}, following the conventions of \cite{Pawlowski:2005xe, Pawlowski:2021tkk}.

From the definition of the metric \labelcref{eq:metric}, we can directly infer the commonly found contractions
\begin{align}
	\gamma^{a}_{\phantom{a}c} &= \gamma^{ab}\gamma_{bc} = (-1)^{ac}\delta^a_c
	\,,\notag\\[1ex]
	\gamma_{a}^{\phantom{a}c} &= \gamma^{bc}\gamma_{ba} = \delta^c_a
	\,,
\end{align}
We briefly note that, as a consequence, $\gamma^{a}_{\phantom{a}c}\gamma^{c}_{\phantom{c}d} = \delta^{a}_d$.

With the metric in hand, any field derivative can be directly expressed using the metric, and one easily derives that
\begin{align}
	\frac{\delta \Phi^b}{\delta \Phi^a} = \gamma_a^{\phantom{a}b}
	\,,\qquad
	\frac{\delta \Phi_b}{\delta \Phi^a} = \gamma_{ab}
	\,,\qquad
	\frac{\delta \Phi^b}{\delta \Phi_a} = \gamma^{ab}
	\,,\qquad
	\frac{\delta \Phi_b}{\delta \Phi_a} = \gamma^{a}_{\phantom{a}b}
	\,.
\end{align}
Similarly, for any correlation function $F$, e.g. the 1PI effective action $F=\Gamma$,
\begin{align}
	\frac{\delta }{\delta \Phi^a} F_{bc\ldots} = F_{abc\ldots}
	\,,\qquad
	\frac{\delta }{\delta \Phi_a} F_{bc\ldots} = F^{a}_{\phantom{a}bc\ldots}
	\,.
\end{align}
Specifically for approaches building on the 1PI effective action, we additionally provide the rules for the relation between the propagator $G^{ab} = \frac{\delta^2 W}{\delta J_a \delta J_b}$ and the 1PI two-point function $\Gamma_{ab} = \frac{\delta^2 \Gamma}{\delta \Phi^a \delta \Phi^b}$. \\[0.5ex]
We derive the relation between these two from the quantum equation of motion, first using left-derivatives,
\begin{align}\label{eq:propInvLeft}
	\frac{\delta \Gamma}{\delta \Phi^a} &= \gamma^d_{\phantom{d}a}J_d
	\,,\notag\\[1ex]\Leftrightarrow
	\frac{\delta \Gamma}{\delta J_b\delta \Phi^a} &= \gamma^b_{\phantom{b}a}
	\,,\notag\\[1ex]\Leftrightarrow
	\frac{\delta \Phi^c}{\delta J_b}\frac{\delta \Gamma}{\delta \Phi^c\delta \Phi^a} &= \gamma^b_{\phantom{b}a}
	\,,\notag\\[1ex]\Leftrightarrow
	G^{bc}\Gamma_{ca} &= \gamma^b_{\phantom{b}a}
	\,.
\end{align}
and, using right-derivatives,
\begin{align}\label{eq:propInvRight}
	\frac{\delta \Gamma}{\delta \Phi^a\delta J_b} &= \gamma_a^{\phantom{a}b}
	\,,\notag\\[1ex]\Leftrightarrow
	\frac{\delta \Gamma}{\delta \Phi^a\delta \Phi^c}\gamma^c_{\phantom{c}d}\left(\Phi^d\frac{\overset{\leftarrow}{\delta}}{\delta J_b}\right) &= \gamma_a^{\phantom{a}b}
	\,,\notag\\[1ex]\Leftrightarrow
	\Gamma_{ac}\gamma^c_{\phantom{c}d}G^{db} = \Gamma_{ad}(-1)^{dd}G^{db} &= \gamma_a^{\phantom{a}b}
	\,.
\end{align}
We can take a derivative $\frac{\delta}{\delta\Phi^f}$ of \labelcref{eq:propInvLeft}, contract the equation with $G$ from the right, and use \labelcref{eq:propInvRight} to obtain the rule
\begin{align}\label{eq:derivG_full}
	\left(\frac{\delta}{\delta \Phi^f}G^{ba}\right)
	&= (-1)(-1)^{bf}(-1)^{dd}\, G^{bc}\Gamma_{cfd}G^{da}\,,
\end{align}
In the above, $(-1)^{bf}$ comes from commuting the derivative past the first index $b$.

Furthermore, one can take also derivatives with respect to functions with multiple superindices, which is used e.g. in the derivation of nPI equations. The derivative for such objects reads
\begin{align}
	\frac{\delta}{\delta R_{a_1a_2\ldots}} R_{b_1b_2\ldots} =
	\textrm{Sym}_{a_1a_2\ldots}\left(
	\gamma^{a_1b_1}\gamma^{a_2b_2}\ldots
	+ \gamma^{a_1b_2}\gamma^{a_2b_1}\ldots
	+ \textrm{all commutations}
	\right)\,,
	\label{eq:symDeriv}
\end{align}
where $\textrm{Sym}_{a_1a_2\ldots}$ is the symmetry factor from arbitrary permutations of identical fields from the given set of indices. For example,
\begin{align}
	\textrm{Sym}_{\phi_1\phi_2\phi_3q_4q_5\bar{q}_5} = \frac{1}{3!}\frac{1}{2!}\frac{1}{1!}\,.
\end{align}
When building a DSE, \FunKit assumes the following form of a classical action:
\begin{align}
	S[\phi] = \sum_{\{i_1\ldots i_n\} \in \text{Trunc}(S)} S_{i_n\ldots i_1} \phi^{i_1}\cdots\phi^{i_n}\,,
\end{align}
where $\text{Trunc}(S)$ represents the set of all (ordered) correlation functions included in the truncation list for $S$, and $\phi$ represents the classical superfield.
As an example, this results in a classical action for the Yang--Mills theory that reads
\begin{align}
	S_\textrm{YM} &= S_{A^{i_2}A^{i_1}}A^{i_1}A^{i_2} + S_{c^{i_2}\bar{c}^{i_1}}\bar{c}^{i_1}c^{i_2}
	\notag\\[0.5ex]
	&\qquad + S_{A^{i_3}A^{i_2}A^{i_1}}A^{i_1}A^{i_2}A^{i_3} + S_{A^{i_4}A^{i_3}A^{i_2}A^{i_1}}A^{i_1}A^{i_2}A^{i_3}A^{i_4} + S_{A^{i_3}c^{i_2}\bar{c}^{i_1}}\bar{c}^{i_1}c^{i_2}A^{i_3}\,.
\end{align}
This is identical with how \QMeS defines its classical action. Furthermore, it is consistent with simply following our convention that
\begin{align}
	S_{i_1\ldots i_n} = \frac{\delta}{\delta\phi^{i_1}}\cdots\frac{\delta}{\delta\phi^{i_n}}S[\phi]\,.
\end{align}
\DoFun~3 \cite{Huber:2019dkb} and \FunKit{} share their stated conventions, with the exception that \DoFun{} extracts the legs of every vertex  with $n\ge 3$ in \emph{reversed} order. Relative to a vertex from \FunKit, \DoFun carries an additional sign, to be precise
\begin{align}
	F_{a_1a_2\ldots a_n}^\textrm{FunKit} &= (-1)^{n_G(n_G-1)/2}\, F_{a_1a_2\ldots a_n}^\textrm{DoFun}\,,\forall n \geq 3\,,
	\notag\\[1ex]
	F_{a_1a_2}^\textrm{FunKit} &= F_{a_1a_2}^\textrm{DoFun}\,,
\end{align}
where $n_G$ is the number of distinct Grassmann fields within $\{\Phi^{a_1},\ldots,\Phi^{a_n}\}$ and $F$ is a correlation function, e.g. $F=\Gamma,S,\ldots$.
\FunKit's \mathem{\mmaDef{FunKitForm}} and \mathem{\mmaDef{DoFunForm}} apply this correction automatically; users porting \DoFun{} code by hand must track it explicitly.

Finally, in any given vertex we always consider all momenta to be incoming, which also fixes our Fourier convention:
\begin{equation}
	\Phi^a(x) = \int \frac{d^d p}{(2\pi)^d} e^{ipx}\Phi^a(p)
	\,.
\end{equation}

%%%%%%%%%%%%%%%%%%%%%%%%%%%%%%
\section{NJL Model Example}
\label{app:NJL}

In this appendix, we provide a complete worked example of defining custom tensor bases and deriving flow equations for an NJL model \cite{Klevansky:1992qe} with Dirac fermions that have no further group structure, described by a classical action at some fixed UV scale $\Lambda$,
\begin{align}
	\Gamma_{\textrm{4F},\Lambda} = \bar\psi (\partial_\mu\gamma_\mu + m)\psi + \sum_{i=1}^3 \lambda_i\mathcal{T}^{(i)}_{d_1d_2d_3d_4}\bar\psi_{d_1}\bar\psi_{d_2}\psi_{d_3}\psi_{d_4}\,.
\end{align}
Here, $\mathcal{T}^{(i)}_{d_1d_2d_3d_4}$ is the $i$-th tensor structure of the full four-fermion interaction and $\lambda_i$ is the corresponding dressing (which has, in general, momentum dependence that we suppress here for brevity).

To obtain flow equations for this theory, as a first step we use the \TensorBases package to create the two-fermion basis and then the four-fermion basis:
\begin{mmaCell}{Input}
\mmaDef{TBConstructBasis}[
 "Name" -> "psibarpsi",
 "Vertex" -> \{psibar, psi\},
 "VertexStructure" -> Tensor[1,2],
 "InnerProduct" -> Tensor1[1,2] Tensor2[2,1],
 "CanonicalProduct" -> Tensor1[1,2] Tensor2[2,1],
 "Indices" -> \{\{p1,d1\}, \{p2,d2\}\},
 "Tensors" -> \{\{deltaDirac[d1,d2], gamma[mu,d1,d2] vec[p1,mu]\}\}];

\mmaDef{TBConstructBasis}[
 "Name" -> "4FBasis",
 "Vertex" -> \{psibar, psi, psibar, psi\},
 "VertexStructure" -> 2 Tensor[1,2,3,4] - 2 Tensor[3,2,1,4],
 "InnerProduct" -> 2 Tensor1[1,2,3,4] (Tensor2[2,1,4,3] - Tensor2[4,1,2,3]),
 "CanonicalProduct" -> Tensor1[1,2,3,4] Tensor2[2,1,4,3],
 "Indices" -> \{\{p1,d1\}, \{p2,d2\}, \{p3,d3\}, \{p4,d4\}\},
 "Tensors" -> \{\{deltaDirac[d1,d2] deltaDirac[d3,d4],
                gamma[mu,d1,d2] gamma[mu,d3,d4],
                I gamma[mu,d1,d1int] gamma5[d1int,d2] I gamma[mu,d3,d3int] gamma5[d3int,d4],
                gamma5[d1,d2] gamma5[d3,d4],
                sigma[mu,nu,d1,d2] sigma[nu,mu,d3,d4]\}\}];
\end{mmaCell}
In the above, we have defined two new bases, one for the fermion propagator \mathem{"psibarpsi"}, and one for the four-fermion interaction \mathem{"4FBasis"}. The meaning of the keys is described in more detail in \cite{Braun:2025gvq} and the \TensorBases documentation. However, we very briefly summarise the most important points:
\begin{itemize}
	\item \mathem{"Name"}: A unique string identifier for the basis, used in all subsequent references.
	\item \mathem{"Vertex"}: List of fields making up the vertex, e.g.\ \mathem{{psibar, psi}} for a two-point function. The order is significant.
	\item \mathem{"VertexStructure"}: How to map a basis element onto a vertex (i.e., taking all derivatives of attached fields). Specified using an expression containing \mathem{\mmaDef{Tensor}[1,2,...]}, where the numbers specify the fields in \mathem{"Vertex"} counted from the first list element.
	\item \mathem{"InnerProduct"}: Defines the trace operation used to compute the metric of the basis, expressed via \mathem{\mmaDef{Tensor1}} and \mathem{\mmaDef{Tensor2}} contracting two basis elements. Usually, this is given by the (coordinate-parametrised) action of the tensor in the dual tangent bundle of the field space manifold on a tensor in the tangent bundle, i.e.,
	\begin{align}
		\langle\mathcal{T}^{(i)},\mathcal{T}^{(j)}\rangle = \mathcal{T}^{(i)}_{a_1a_2\ldots}
		\frac{\delta}{\delta\Phi^{a_1}}\frac{\delta}{\delta\Phi^{a_2}}\cdots
		\mathcal{T}^{(j)}_{b_1b_2\ldots}\Phi^{b_1}\Phi^{b_2}\cdots\,.
	\end{align}
	\item \mathem{"CanonicalProduct"}: A simplified contraction pattern (same syntax as \mathem{"InnerProduct"}) used for projections.
	\item \mathem{"Indices"}: A list of index sets for each leg, specifying here the momentum and internal (e.g.\ Dirac) indices, e.g.\ \mathem{{{p1,d1},{p2,d2}}}.
	\item \mathem{"Tensors"}: The candidate tensor structures from which the basis is constructed. Each sub-list forms a tensor space; the function takes the full tensor product first and then reduces the combined set to a maximal linearly independent subset using the given inner product.
\end{itemize}
With our bases now known to \TensorBases, we define the corresponding setup, which has only two-fermion and four-fermion interactions:
\begin{mmaCell}{Input}
fields4F = <|
   "Commuting" -> \{\},
   "Grassmann" -> \{\{psibar[p,\{d\}], psi[p,\{d\}]\}\}
  |>;
truncation4F = <|
   \mmaDef{GammaN} -> \{\{psibar,psi,psibar,psi\}, \{psibar,psi\}\},
   \mmaDef{Propagator} -> \{\{psibar,psi\}\},
   \mmaDef{Rdot} -> \{\{psibar,psi\}\}
  |>;
bases4F = <|
   \mmaDef{GammaN} -> \{
     \{psibar,psi,psibar,psi\} -> "4FBasis",
     \{psibar,psi\} -> "psibarpsi"\},
   \mmaDef{Propagator} -> \{\{psibar,psi\} -> "psibarpsi"\},
   \mmaDef{Rdot} -> \{\{psibar,psi\} -> \{"psibarpsi",2\}\}
  |>;
setup4F = <|
   "FieldSpace" -> fields4F,
   "Truncation" -> truncation4F,
   "FeynmanRules" -> bases4F
  |>;
\end{mmaCell}
Note that we do not set the above defined setup as the global setup, and thus explicitly specify it in all function calls in this appendix.
Now, we can already create the diagrammatic rules for our setup:
\begin{mmaCell}{Input}
\mmaDef{Propagator}[\{psi,psibar\},\{\{p1,\{d1\}\},\{p2,\{d2\}\}\}]/.\mmaDef{FMakeDiagrammaticRules}[\mmaDef{setup4F}]
\end{mmaCell}
\begin{mmaCell}{Output}
\mmaFrac{-deltaDirac[d1,d2] dressing[InverseProp,\{psibar,psi\},1,\{p1,p2\}]}{\mmaSup{dressing[InverseProp,\{psibar,psi\},1,\{p1,p2\}]}{2}-\mmaSup{dressing[InverseProp,\{psibar,psi
\},2,\{p1,p2\}]}{2} sp[p1,p1]}+
\mmaFrac{dressing[InverseProp,\{psibar,psi\},2,\{p1,p2\}] gamma[mu$24889,d1,d2] vec[p1,mu$24889]}{-\mmaSup{dressing[InverseProp,\{psibar,psi\},1,\{p1,p2\}]}{2}+\mmaSup{dressing[InverseProp,\{psibar,psi\},2,\{p1,p2\}]}{2} sp[p1,p1]}
\end{mmaCell}
We see again that the result of \mathem{\mmaDef{FMakeDiagrammaticRules}[\mmaDef{setup4F}]} is a replacement table that inserts all tensors in the specified basis for a given object, here the propagator. Each tensor is multiplied with a momentum-dependent \mathem{\mmaDef{dressing}}, as seen above.

This form still requires the user to specify a parametrisation for the dressings. Therefore, we also introduce a parametrisation for the dressing functions that feature in the diagrammatic rules,
\begin{mmaCell}{Input}
parametrization4F = \{
  \mmaDef{dressing}[GammaN, \{psibar, psibar, psi, psi\}, i_, \{__\}] :> lambda[i],
  \mmaDef{dressing}[Rdot, \{psibar, psi\}, 2, \{lf1, -lf1\}] :> Rdot[lf1],
  \mmaDef{dressing}[InverseProp, \{psibar, psi\}, i_, \{p1_, p2_\}] :> GInv[i, p1]
 \};
\end{mmaCell}
To keep the output concise, we have assumed momentum-independent dressings, explicitly \mathem{lambda[i]} for the \mathem{i}-th vertex.
With all the above, the two-point function reads
\begin{mmaCell}{Input}
Flow2F = \mmaDef{FTakeDerivatives}[\mmaDef{setup4F},\mmaDef{WetterichEquation},\{psibar[i1],psi[i2]\}];
Flow2FTrunc = \mmaDef{FTruncate}[\mmaDef{setup4F}, Flow2F];
\mmaDef{FPlot}[\mmaDef{setup4F}, Flow2FTrunc];
\end{mmaCell}
\vspace{-1ex}

\noindent\hspace{3.0em}\includegraphics[width=0.11\linewidth]{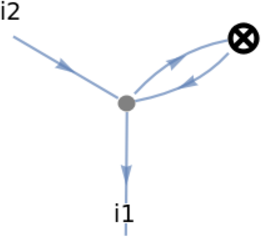}\\
After routing and tracing, we obtain the flow equation as
\begin{mmaCell}{Input}
Flow2FRouted = \mmaDef{FRoute}[\mmaDef{setup4F}, \mmaDef{Flow2FTrunc}];
traceExpr2F = Flow2FRouted["1-Loop"]["Expression"]/.\mmaDef{FMakeDiagrammaticRules}[\mmaDef{setup4F}];
proj2F = \mmaDef{TBGetProjector}["psibarpsi", 1, \{i2,i1\}/.Flow2FRouted["1-Loop"]["ExternalIndices"]];
Flow2FTraced = \mmaDef{FTerm}[proj2F, traceExpr2F ]//\mmaDef{FormTrace};
Flow2FTraced/.\mmaDef{parametrization4F}//\mmaDef{Simplify}
\end{mmaCell}
\begin{mmaCell}{Output}
\{\mmaFrac{4 GInv[1,lf1] GInv[2,lf1] (3 lambda[1] - 4(lambda[2] + lambda[3])) Rdot[lf1] sp[lf1,lf1]}{\mmaSup{(\mmaSup{GInv[1,lf1]}{2} - \mmaSup{GInv[2,lf1]}{2} sp[lf1,lf1])}{2}}\}
\end{mmaCell}
Note here that we had to swap the fermion and anti-fermion in the projector in order to correctly route the Dirac tensor indices. One should be mindful of the additional minus arising from the swapping of fields.

In the same way, we can derive the flow of the first tensor structure of the four-fermion basis:
\begin{mmaCell}{Input}
Flow4F = \mmaDef{FTakeDerivatives}[\mmaDef{setup4F},\mmaDef{WetterichEquation},\{psibar[i1],psi[i2],psibar[i3],psi[i4]\}];
Flow4FTrunc = \mmaDef{FTruncate}[\mmaDef{setup4F},Flow4F];
\mmaDef{FPlot}[\mmaDef{setup4F},Flow4FTrunc]
\end{mmaCell}

\noindent\hspace{3.0em}\includegraphics[width=0.7\linewidth]{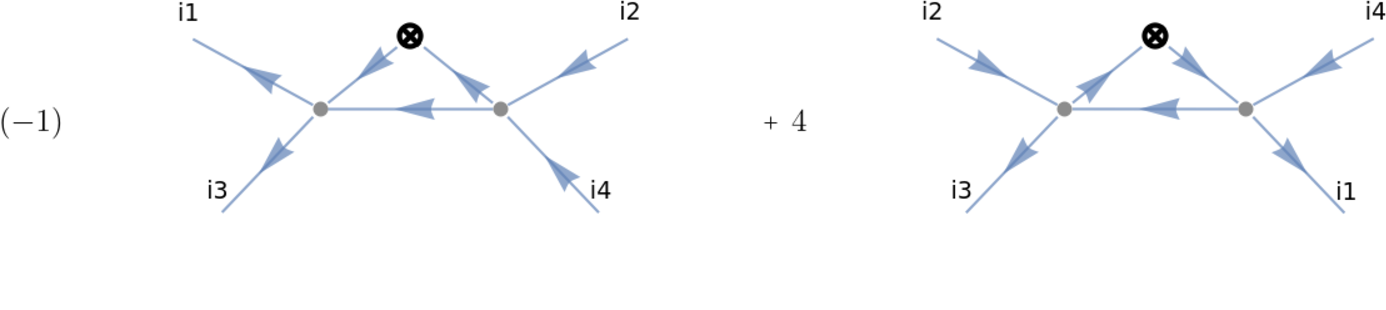}

\begin{mmaCell}{Input}
Flow4FRouted = \mmaDef{FRoute}[\mmaDef{setup4F},\mmaDef{Flow4FTrunc}];
traceExpr4F = Flow4FRouted["1-Loop"]["Expression"]/.\mmaDef{FMakeDiagrammaticRules}[\mmaDef{setup4F}];
proj4F = \mmaDef{TBGetProjector}["4FBasis",1,\{i2,i1,i4,i3\}/.Flow4FRouted["1-Loop"]["ExternalIndices"]];
Flow4FTraced = FTerm[proj4F,traceExpr4F]//\mmaDef{FormTrace};
Flow4FTraced/.\mmaDef{parametrization4F}//\mmaDef{Simplify}
\end{mmaCell}
\begin{mmaCell}{Output}
\{(8 Rdot[lf1] (4 GInv[1,lf1] GInv[1,lf1-p2-p3] GInv[2,lf1] (5 lambda[1]^2-12 lambda[1] (lambda[2]+lambda[3])+8 (3 lambda[2]^2-2 lambda[2] lambda[3]+3 lambda[3]^2)) sp[lf1,lf1]
    +GInv[2,lf1-p2-p3] (7 lambda[1]^2+32 lambda[3] (lambda[2]+lambda[3])-4 lambda[1] (3 lambda[2]+7 lambda[3])) (GInv[1,lf1]^2+GInv[2,lf1]^2 sp[lf1,lf1]) (sp[lf1,lf1]-sp[lf1,p2]-sp[lf1,p3])))/((GInv[1,lf1]^2-GInv[2,lf1]^2 sp[lf1,lf1])^2 (-GInv[1,lf1-p2-p3]^2+GInv[2,lf1-p2-p3]^2 
    (sp[lf1,lf1]-2 sp[lf1,p2]-2 sp[lf1,p3]+sp[p2,p2]+2 sp[p2,p3]+sp[p3,p3]))),(8 Rdot[lf1] (GInv[1,lf1] GInv[1,-lf1+p1+p3] GInv[2,lf1] (lambda[1]^2+16 (lambda[2]-lambda[3])^2) sp[lf1,lf1]
    -4 GInv[2,-lf1+p1+p3] lambda[1] (lambda[2]-lambda[3]) (GInv[1,lf1]^2+GInv[2,lf1]^2 sp[lf1,lf1]) (sp[lf1,lf1]-sp[lf1,p1]-sp[lf1,p3])))/((GInv[1,lf1]^2-GInv[2,lf1]^2 sp[lf1,lf1])^2 (-GInv[1,-lf1+p1+p3]^2+GInv[2,-lf1+p1+p3]^2 (sp[lf1,lf1]-2 sp[lf1,p1]-2 sp[lf1,p3]+sp[p1,p1]+2 
    sp[p1,p3]+sp[p3,p3])))\}
\end{mmaCell}
Just like in the two-point example above, we had to swap fermion and anti-fermion indices for a consistent chain of Dirac indices.

%%%%%%%%%%%%%%%%%%%%%%%%%%%%%%
\section{Optimisation of powers for C++}
\label{app:powr}

\FunKit replaces integer powers of expressions in C++ with the function \cpp{powr<n>(expr)}. This allows for a more efficient implementation of (low) integer powers than the standard \cpp{std::pow(expr,n)}. For convenience, a possible implementation of \cpp{powr} may read
\begin{lstlisting}[language=c++,style=myStyle]
template <int n, typename NumberType> requires requires(NumberType x) {
  x * x;
  static_cast<NumberType>(1) / x;
}
constexpr NumberType powr(const NumberType x)
{
  if constexpr (n == 0) return static_cast<NumberType>(1);
  else if constexpr (n < 0) return static_cast<NumberType>(1)/powr<-n,NumberType>(x);
  else if constexpr (n == 1) return x;
  else if constexpr (n % 2 == 0) return powr<n/2>(x) * powr<n/2>(x);
  else return powr<n/2>(x) * powr<n/2>(x) * x;
}
\end{lstlisting}
One can turn off usage of \cpp{powr} by setting
\begin{mmaCell}{Input}
\mmaDef{FUseCppPowr}[\mmaDef{False}];
\end{mmaCell}
%

%%%%%%%%%%%%%%%%%%%%%%%%%%%%%%
\section{Structure of the Package}
\label{app:Structure}

\begin{figure*}[t]
	\begin{tikzpicture}[node distance=2.5cm, auto,>=latex']
		% Nodes
		\node [block] (AnSEL) {\AnSEL \\(simplification, routing)};
		\node [block, below of=AnSEL] (DiANE) {\DiANE (visualisation)};
		\node [block, below of=DiANE, xshift=15ex, yshift=7ex] (COEN) {\COEN (code output)};
		
		\node [block, right of=AnSEL, yshift=-8ex, xshift=12ex] (FEDeriK) {\FEDeriK (syntax)};
		
		\node [block, right of=FEDeriK, node distance=5cm, xshift=0ex] (TRACY) {\TRACY (tracing)};
		\node [block, above of=TRACY] (TensorBases) {\TensorBases};
		\node [block, below of=TRACY, xshift=-15ex, yshift=5ex] (DiRK) {\DiRK \\(diagrammatic rules)};

		% Arrows
		\draw [->] (FEDeriK) -- (TRACY);
		
		\draw [->] (TensorBases) -- (TRACY);
		\draw [->] (FEDeriK) -- (AnSEL);
		\draw [->] (FEDeriK) -- (DiRK);
		\draw [->] (FEDeriK) -- (TRACY);
		\draw [->] (FEDeriK) -- (DiANE);
		\draw [->] (AnSEL) -- (DiANE);
		\draw [->] (FEDeriK) -- (COEN);
	\end{tikzpicture}
	
	\caption{Dependency structure of the submodules of \FunKit. Note that this does not reflect the user workflow one-to-one, see \Cref{fig:workflow}.
	}
	\label{fig:FunKit_structure}
\end{figure*}
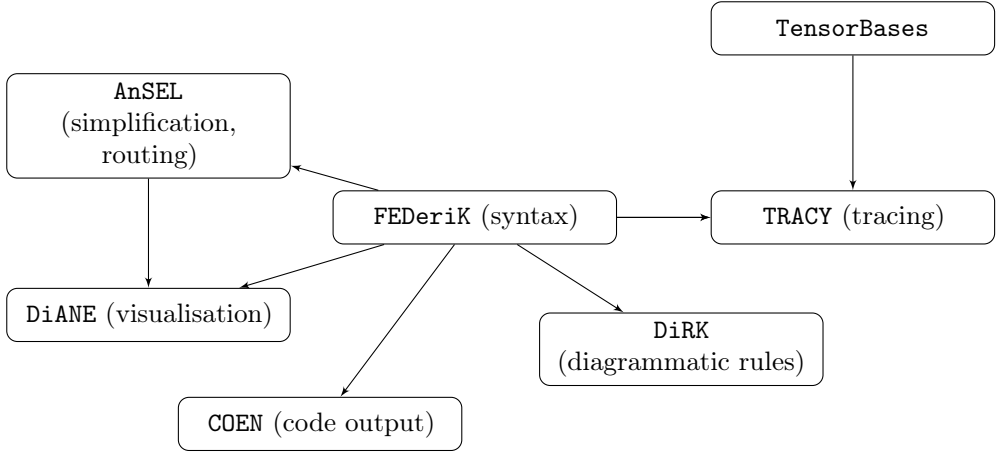

Both for maintainability and usability, \FunKit is structured into modules, which partially depend on each other. A dependency graph of the modules is shown in \Cref{fig:FunKit_structure}. 
When \FunKit is imported for the first time, all modules are automatically loaded in the correct order. The modules are:
\begin{itemize}
	\item \FEDeriK, the {\textbf{F}unctional \textbf{E}quation \textbf{Deri}vation \textbf{K}it} is the core module of \FunKit, providing all the logic to derive functional equations. It provides the base syntax for the description of functional equations, gives facilities to perform functional derivatives and series expansions of derivative operators and fields.
	\item \AnSEL: The \textbf{An}alysis and \textbf{S}implification of \textbf{E}quations with \textbf{L}oops module, which provides tools to group identical diagrams and for momentum- and index-routing in arbitrary multi-loop diagrams.
	\item \DiRK: The \textbf{Di}agrammatic \textbf{R}ule \textbf{K}it, containing tools to create Feynman rules from a given set of tensorial bases.
	\item \TensorBases: An external package to define tensor bases for correlation functions, see \cite{Braun:2025gvq}. This package will be automatically installed with FunKit, either during the installation, or when one imports \FunKit for the first time in Mathematica.
	\item \TRACY gives a convenient interface to \FormTracer \cite{Cyrol:2016zqb} to trace out tensor structures in diagrams derived with \FunKit. \FormTracer is an external package which is automatically installed by \FunKit.
	\item \DiANE: The \textbf{Di}agram \textbf{A}rts and \textbf{N}otation of \textbf{E}quations module, which provides tools to visualise diagrams and export \LaTeX{} code.
	\item \COEN: The \textbf{CO}de \textbf{EN}gine module, which provides tools to generate computer code in C++, Julia or Fortran from traced diagrams for further processing in a numerical framework.
\end{itemize}
The modules are located in the repository folder \texttt{FunKit/modules/}.
Each module is split into files containing one or a few functions each in an appropriate subfolder, e.g. \texttt{FunKit/modules/FEDeriK/Notation.m} which contains all rules pertaining to \mathem{\mmaDef{FTerm}} and \mathem{\mmaDef{FEx}}.

% Flowchart of the structure

%%%%%%%%%%%%%%%%%%%%%%%%%%%%%%
\section{Testing and Verification}
\label{app:algo_test}

The correctness of the output given by \FunKit is verified by a suite of unit tests that cover all of the most important methods provided by \FunKit.
These can be run either directly from a Mathematica notebook by using the command
\begin{mmaCell}{Input}
\mmaDef{FTest}[]
\end{mmaCell}
or alternatively by using the \texttt{CMake} integration of \FunKit from the top directory of the repository:
\begin{lstlisting}[language=Bash]
mkdir build
cd build
cmake ..
make test
\end{lstlisting}
A single test can be run by
\begin{lstlisting}[language=Bash]
make test-single FILE=FEDeriK/FunctionalDTests.m
\end{lstlisting}
In addition to unit tests, we also verify that \FunKit produces the same diagrams (and signs) as \QMeS and \DoFun. 
Exhaustive tests are run for one- to four-point functions (using both fRG and DSE master equations) in the case of a purely scalar, a Yukawa theory, and a Yang--Mills theory, see \texttt{tests/CrossTests/}.

%%%%%%%%%%%%%%%%%%%%%%%%%%%%%%
\section{Overview of all functions in \FunKit}
\label{app:all_func}

All of the listed functions that take a \mathem{setup} as the first argument can be also called without it, if a global setup has been set beforehand using \mathem{\mmaDef{FSetGlobalSetup}[setup]}.

Besides the explanations in this document, \FunKit gives detailed documentation on each function in its usage string, i.e., an expression like \mathem{\mmaDef{FTakeDerivatives}::\mmaCmt{usage}} will give explanations on how to use the function.

\begin{tcolorbox}[title=Master Equations,
	title filled=false,
	colback=pastelblue!5!white,
	colframe=pastelblue,
	fonttitle=\bfseries]
	\begin{tabular}{p{4.6cm} p{10.7cm}}
		\mathem{\mmaDef{FMakeClassicalAction}[setup]} & 
		\small Constructs the classical action from the setup's truncation table for the object \mathem{\mmaDef{S}}.
		\\[3ex]
		\mathem{\mmaDef{FMakeDSE}[setup, field]} & 
		\small Constructs the Dyson--Schwinger equation $\Gamma_{b} = \left\langle \frac{\delta S[\phi]}{\delta \phi^b} \right\rangle_{\phi^a = \Phi^a + G^{ab}\frac{\delta}{\delta\Phi^b}}$.
		\\[3ex]
		\mathem{\mmaDef{WetterichEquation}} & 
		\small The Wetterich equation for the functional Renormalisation Group, ${\partial_t \Gamma = (1/2) G^{ab} (\partial_t R)_{ab}}$.
		\\[3ex]
		\mathem{\mmaDef{GeneralizedFlowEquation}} & 
		\small The generalised fRG flow equation with flowing field reparametrisation $\dot\Phi$, ${\partial_t \Gamma = -\dot\Phi^a\Gamma_a + \frac{1}{2}G^{ac}(\gamma_c^{\phantom{c}b}\partial_t + 2\frac{\delta\dot{\Phi}^b}{\delta\Phi^c}){R}_{ab}}$.
		\\[3ex]
		\mathem{\mmaDef{FEmptySetup}} & 
		\small Not a master equation, but an empty setup with no fields or truncation tables defined. Useful for deriving fully abstract equations.
	\end{tabular}
\end{tcolorbox}

\begin{tcolorbox}[title=Diagrams and \LaTeX,
title filled=false,
colback=pastelblue!5!white,
colframe=pastelblue,
fonttitle=\bfseries]
\begin{tabular}{p{4.2cm} p{11.0cm}}
	\mathem{\mmaDef{FPrint}[setup, expr]} & 
	\small Prints functional expressions in formatted mathematical notation using \LaTeX{} rendering with the \texttt{MaTeX} package. Returns the original expression after printing for easy chaining.
	\\[3ex]
	\mathem{\mmaDef{FTex}[setup, expr]} & 
	\small Outputs a string of \LaTeX{} code for the given expression.
	\\[3ex]
	\mathem{\mmaDef{FPlot}[setup, expr]} & 
	\small Prints Feynman Diagrams of the given functional expression.
	\\[3ex]
	\mathem{\mmaDef{FAddTexStyles}[rules...]} & 
	\small Adds custom \LaTeX{} styling rules for specific mathematical objects or fields. Style rules should be given as replacement rules, e.g., \mathem{field -> "\\mathbf{field}"}.
	\\[3ex]
	\mathem{\mmaDef{FSetTexStyles}[rules...]} & 
	\small Same as \mathem{\mmaDef{FAddTexStyles}}, but resets the list of known style rules.
\end{tabular}
\end{tcolorbox}

\begin{tcolorbox}[title=Notation,
	title filled=false,
	colback=pastelblue!5!white,
	colframe=pastelblue,
	fonttitle=\bfseries]
	\begin{tabular}{p{3cm} p{12.2cm}}
		\mathem{\mmaDef{FEx}} & 
		\small Represents a sum of \mathem{\mmaDef{FTerm}} objects and is the main container for functional equations in \FunKit. Supports non-commutative multiplication \mathem{**} and handles some simplifications.
		\\[3ex]
		\mathem{\mmaDef{FTerm}} &
		\small A single term in a functional equation, i.e., a product of factors. Factors can be numbers, objects, fields, derivative operators, or appropriate combinations thereof. Putting two Grassmann fields into the same factor will lead to errors being thrown.
		\\[3ex]
		\mathem{\mmaDef{Propagator}} &
		\small Two-index object representing the propagator $G^{ab}$.
		\\[3ex]
		\mathem{\mmaDef{GammaN}} &
		\small Indexed object representing the n-point 1PI correlation function $\Gamma_{ab\ldots}$.
		\\[3ex]
		\mathem{\mmaDef{S}} &
		\small Indexed object representing the classical n-point vertex $S_{ab\ldots}$.
		\\[3ex]
		\mathem{\mmaDef{R}} &
		\small Two-index object representing the regulator $R_{ab}$.
		\\[3ex]
		\mathem{\mmaDef{Rdot}} &
		\small Two-index object representing the regulator derivative $\partial_tR_{ab}$.
		\\[3ex]
		\mathem{\mmaDef{Phidot}} &
		\small Indexed object representing (derivatives of) the field reparametrisation $\dot\Phi^{a}$
		\\[3ex]
		$\gamma$ &
		\small Two-index object representing the field metric $\gamma^{ab}$.
		\\[3ex]
		\mathem{\mmaDef{Field}} &
		\small Any field can either be written as \mathem{A[i]}, or as \mathem{\mmaDef{Field}[{A},{i}]}.
		\\[3ex]
		\mathem{\mmaDef{FMinus}} &
		\small Two-index object representing the sign function $(-1)^{ab}$.
		\\[3ex]
		\mathem{\mmaDef{SymmetryFactor}} &
		\small Indexed object representing the symmetry factor $\textrm{Sym}_{ab\ldots}$.
		\\[3ex]
		\mathem{\mmaDef{AnyField}} &
		\small A tag representing a yet unspecified field.
		\\[3ex]
		\mathem{\mmaDef{FDOp}} &
		\small A derivative operator, taking a single field or correlation function, e.g., \mathem{\mmaDef{FDOp}[A[i1]]} or  \mathem{\mmaDef{FDOp}[\mmaDef{R}[{\mmaDef{AnyField},\mmaDef{AnyField}},{i1,i2}]]}
		\\[3ex]
		\mathem{\mmaDef{dressing}} &
		\small Represents the diagrammatic dressing of objects with specific field content. The full form is
		\mathem{\mmaDef{dressing}[object,{f1,f2,...},index,{p1,p2,...}]}, where \mathem{{f1,f2,...}} specifies the fields, \mathem{index} the basis element the dressing pertains to, and \mathem{{p1,p2,...}} are the momenta of the fields.
		\\[3ex]
		\mathem{\mmaDef{InverseProp}} &
		\small Special object, with notation as for indexed objects, that represents the full inverse propagator $(G^{-1})^{ab} = \Gamma^{ab} + \textrm{other}$, where additional terms may arise due to a modification of the Legendre transformation, as usually done in an fRG context. 
	\end{tabular}
\end{tcolorbox}

\begin{tcolorbox}[title=Compatibility,
	title filled=false,
	colback=pastelblue!5!white,
	colframe=pastelblue,
	fonttitle=\bfseries]
	\begin{tabular}{p{4cm} p{11.2cm}}
		\mathem{\mmaDef{QMeSForm}[setup,expr]} & 
		\small Converts expressions containing indexed objects (like \mathem{\mmaDef{Propagator}}, \mathem{\mmaDef{GammaN}}) to \QMeS-style notation. Uses the canonical ordering \mathem{"b>af>f"} for field arrangement.
		\\[3ex]
		\mathem{\mmaDef{DoFunForm}[setup,expr]} & 
		\small Converts expressions containing indexed objects (like \mathem{\mmaDef{Propagator}}, \mathem{\mmaDef{GammaN}}) to \DoFun-style notation.
		\\[3ex]
		\mathem{\mmaDef{FunKitForm}[setup,expr]} & 
		\small Transforms expressions from \QMeS-style or \DoFun-style notation to \FunKit-style notation.
	\end{tabular}
\end{tcolorbox}

\begin{tcolorbox}[title=Derivatives and truncations,
	title filled=false,
	colback=pastelblue!5!white,
	colframe=pastelblue,
	fonttitle=\bfseries]
	\begin{tabular}{p{5.7cm} p{9.5cm}}
		\mathem{\mmaDef{FTakeDerivatives}[setup,expr,dList]} &
		\small Takes multiple functional derivatives of \mathem{expr} with respect to the fields in \mathem{dList}, starting from the right to the left.
		Returns the result with all derivative operators resolved.\newline
		\mathem{dList} must be a list of fields like \mathem{{A[i1],A[i2]}}.
		\\[3ex]
		\mathem{\mmaDef{FResolveDerivatives}[setup,expr]} &
		\small Iteratively resolves all functional derivative operators \mathem{\mmaDef{FDOp}} in the expression.
		\\[3ex]
		\mathem{\mmaDef{FResolveFDOp}[setup,expr]} &
		\small Resolves the right-most \mathem{\mmaDef{FDOp}} in every \mathem{\mmaDef{FTerm}}.
		\\[3ex]
		\mathem{\mmaDef{FTruncate}[setup,expr]} &
		\small Truncates \mathem{expr} according to the truncation table in \mathem{setup}. Replaces undetermined \mathem{\mmaDef{AnyField}} with explicit fields and removes terms not included in the truncation.
		Open indices are ignored and the expression must not contain unresolved \mathem{\mmaDef{FDOp}}.
		\\[3ex]
		\mathem{\mmaDef{FTruncateOpenIndices}[setup,expr]} &
		\small Truncates open indices in \mathem{expr} according to the truncation tables specified in the \mathem{setup}.
		Explicitly, a sum is taken over all possible fields for every single open-index \mathem{\mmaDef{AnyField}} in the expression.
		\\[3ex]
		\mathem{\mmaDef{FMakeSymmetryList}[setup,fields]} &
		\small Returns a list of symmetries between the provided fields with indices. \mathem{fields} must be given as a list of indexed fields, e.g., \mathem{{f1[i1],...}}.\newline
		Any \mathem{\mmaDef{FEx}} can be annotated as \mathem{\mmaDef{FEx}[..., "Symmetries"->symmetryList]}.
		\\[3ex]
		\mathem{\mmaDef{FSimplify}[setup,expr]} &
		\small Simplifies \mathem{expr} by identifying and combining terms. Optionally, one can either pass a symmetry list directly \mathem{"Symmetries"->symmetryList}, or include it in the given \mathem{\mmaDef{FEx}} - both will be taken into account.
		 Disconnected diagrams are simplified component-wise; reorderings of disconnected components are recognised as equivalent and merged with the correct sign from permuting the sub-graphs.
		\\[3ex]
		\mathem{\mmaDef{FOrderFields}[setup,expr]} &
		\small Orders all fields within \mathem{expr} according to the canonical ordering currently set for FunKit.
		Normally, FunKit takes care of field ordering automatically, but this function allows manual reordering when needed.
		\\[3ex]
		\mathem{\mmaDef{FExpand}[setup,expr,order]} &
		\small Expands powers of FTerm and FEx expressions up to the specified order. This is useful for expanding expressions like $(FTerm[...])^n$ into explicit sums of terms. Automatically fixes indices to ensure uniqueness after expansion.
		\\[3ex]
		\mathem{\mmaDef{DExpand}[setup,expr,order]} &
		\small Expands expressions containing \mathem{\mmaDef{FDOp}} up to the specified order. Similar to \mathem{\mmaDef{FExpand}}, but specifically handles expansions involving functional derivatives.
		\\[3ex]
		\mathem{\mmaDef{FDisconnectedQ}[setup,expr]} &
		\small Checks whether a given \mathem{\mmaDef{FTerm}} represents a connected or disconnected diagram. Returns \mathem{\mmaDef{True}} if it is disconnected.
	\end{tabular}
\end{tcolorbox}

\begin{tcolorbox}[title=Routing,
	title filled=false,
	colback=pastelblue!5!white,
	colframe=pastelblue,
	fonttitle=\bfseries]
	\begin{tabular}{p{5.2cm} p{10.0cm}}
		\mathem{\mmaDef{FRoute}[setup, expr]} & 
		\small Routes indices and momenta in functional expressions, organising terms by loop order. For \mathem{\mmaDef{FTerm}}, returns an Association with keys \mathem{"Expression"}, \mathem{"ExternalIndices"}, and \mathem{"LoopMomenta"}.\newline
		For \mathem{\mmaDef{FEx}}, returns an Association with keys like \mathem{"0-Loop"}, \mathem{"1-Loop"}, etc.
		\\[3ex]
		\mathem{\mmaDef{FSetLoopMomentumName}[name]} &
		\small Sets the symbol for loop momentum variables. Must be a string (Default: \mathem{"l"}) which will then lead to momenta \mathem{l1}, \mathem{lf1}, etc.
		\\[3ex]
		\mathem{\mmaDef{FSetRoutingAlgorithm}[algorithm]} &
		\small Selects how \mathem{\mmaDef{FRoute}} handles the routing of momenta. Available choices for \mathem{algorithm} are \mathem{"Default"} (routes momenta such that fermionic momenta flow through fermionic lines) and \mathem{"Regulator"} (never routes momenta through regulators). Note that the second choice can be problematic at finite temperatures, as Matsubara frequencies are not propagated correctly.
		\\[3ex]
		\mathem{\mmaDef{FUnroute}[setup, expr]} &
		\small \textbf{Experimental:} Reverses the index and momentum routing performed by FRoute, i.e., converts routed expressions back to superindex notation.
	\end{tabular}
\end{tcolorbox}

\begin{tcolorbox}[title=Modifying the Notation,
title filled=false,
colback=pastelblue!5!white,
colframe=pastelblue,
fonttitle=\bfseries]
\begin{tabular}{p{5.4cm} p{9.8cm}}
	\mathem{\mmaDef{FAddCorrelationFunction}[obj]} & 
	\small Adds an object that reacts to reordering and to functional derivatives.
	\\[3ex]
	\mathem{\mmaDef{FAddOrderedObject}[obj]} & 
	\small Adds an object that reacts to reordering but not to functional derivatives.
	\\[3ex]
	\mathem{\mmaDef{FAddIndexedObject}[obj]} & 
	\small Adds an object that does not react to reordering or to functional derivatives.
	\\[3ex]
	\mathem{\mmaDef{FAddObject}[obj]} & 
	\small Adds an object that does not react to reordering or to functional derivatives and ignores whether indices are already contracted.
	\\[3ex]
	\mathem{\mmaDef{FShowCorrelationFunctions}[]} & 
	\small Shows all objects that react to reordering and functional derivatives.
	\\[3ex]
	\mathem{\mmaDef{FShowOrderedObjects}[]} & 
	\small Shows all objects that react to reordering.
	\\[3ex]
	\mathem{\mmaDef{FShowIndexedObjects}[]} & 
	\small Shows all objects that require consistently closed indices.
	\\[3ex]
	\mathem{\mmaDef{FShowObjects}[]} & 
	\small Shows all known objects.
	\\[3ex]
	\mathem{\mmaDef{FSetUnorderedIndices}[obj,ind]} &
	\small Specifies which indices of an indexed object should never be reordered during field ordering operations. \mathem{ind} can be either a single index position, or a list thereof. Counting starts from the right at 1.
	\\[3ex]
	\mathem{\mmaDef{FSetSymmetricObject}[obj,fields]} & 
	\small Sets an object with fields \mathem{fields={f1,...}} to be symmetric in all its indices.
	\\[3ex]
	\mathem{\mmaDef{FAddFDRule}[object, wrt, result]} & 
	\small Adds a custom functional derivative rule to be used with priority when taking derivatives. Patterns should be used in general.\newline
	The rule applies to \mathem{object} (e.g., \mathem{\mmaDef{GammaN}[{f__},{i__}]}) when taking a derivative with respect to \mathem{wrt} (e.g., \mathem{ob[{f1_,f2_},{i1_,i2_}]}), which yields the \mathem{result} (e.g., \mathem{MyResult[{f1,f2,f},{i1,i2,i}]}).
	\\[3ex]
	\mathem{\mmaDef{FClearFDRules}[]} & 
	\small Clears all custom functional derivative rules added with \mathem{\mmaDef{FAddFDRule}}.
\end{tabular}
\end{tcolorbox}

\begin{tcolorbox}[title=Diagrammatic rules and Tracing with FORM,
	title filled=false,
	colback=pastelblue!5!white,
	colframe=pastelblue,
	fonttitle=\bfseries]
	\begin{tabular}{p{5.7cm} p{9.6cm}}
		\mathem{\mmaDef{FMakeDiagrammaticRules}[setup]} & 
		\small Generates a list of replacement rules that convert functional expressions into an explicit form that can be traced. The bases must be known to the \TensorBases package beforehand.
		\\[3ex]
		\mathem{\mmaDef{FSetSymmetricDressing}[obj,}
		\newline\phantom{.}\hspace{4cm}\mathem{{f1,...}]} & 
		\small Sets symmetry properties for diagrammatic dressing of objects.
		The order of field arguments is taken to be irrelevant.
		Alternatively, \mathem{\mmaDef{FSetSymmetricDressing}[obj, {f1,f2,...}, {1,3,...}]} allows specifying which indices should be symmetrised.
		\\[3ex]
		\mathem{\mmaDef{FormTrace}[expr]} & 
		\small We extend the \mathem{\mmaDef{FormTrace}} command from \cite{Cyrol:2016zqb} to take \mathem{\mmaDef{FEx}} and \mathem{\mmaDef{FTerm}}. These are automatically traced in parallel, if possible.\newline 
		\mathem{\mmaDef{FormTrace}[name,expr]} caches the result in a folder inside the trace cache called \mathem{name}. \newline
		\mathem{\mmaDef{FormTrace}[expr,preReplRules,postReplRules]} and \mathem{\mmaDef{FormTrace}[name,expr,preReplRules,postReplRules]} allow for specifying  custom \FORM preRepl and postRepl rules, see \cite{Cyrol:2016zqb} and \mathem{\mmaDef{FormTrace}::\mmaCmt{usage}}.
		\\[3ex]
		\mathem{\mmaDef{FFormSimplify}[expr]} &
		\small Simplifies expressions using FORM's output optimisation (O4) algorithms.
		\mathem{\mmaDef{FFormSimplify}[expr,preReplRules,postReplRules]} allows specifying custom \FORM preRepl and postRepl rules, see \cite{Cyrol:2016zqb}.
		\\[3ex]
		\mathem{\mmaDef{FMakeSPFormRule}[{l1,...},p,}
		\newline\phantom{.}\hspace{4cm}\mathem{{p1,...}]} & 
		\small Creates a \FORM rule for a symmetric point momentum configuration.
		The first argument is a list of loop momenta \mathem{{l1,l2,...}}.
		The second argument \mathem{p} is the average momentum scale.
		The third argument is a list of external leg momenta \mathem{{p1,p2,...}}. The rule projects all momenta to a symmetric configuration with average momentum \mathem{p}.
		\\[3ex]
		\mathem{\mmaDef{FMakeSPFiniteTFormRule}[{l1,...},}
		\newline\phantom{.}\hspace{3.7cm}\mathem{p,{p1,...}]} & 
		\small Creates a \FORM rule for a $(d-1)$-dimensional symmetric point configuration, i.e., a purely spatial symmetric point. Parameters are identical to the ones in \mathem{\mmaDef{FMakeSPFormRule}}.
		\\[3ex]
		\mathem{\mmaDef{FMakeP0Rule}[{p1,...},{prj1,...}]} & 
		\small Creates a replacement rule to project temporal components of momentum expressions. The first argument is a list of external leg momenta \mathem{{p1,p2,...}}.
		The second argument is a list \mathem{{prj1,prj2,...}} of values to be projected on for the temporal components of the given momenta.
		I.e., this sets \mathem{vec[pi,0]} to the value of \mathem{prji}.
		\\[3ex]
		\mathem{\mmaDef{FMakeP0FormRule}[{p1,...},}
		\newline\phantom{.}\hspace{3.75cm}\mathem{{prj1,...}]} &  
		\small Creates a \FORM rule to project temporal components of momentum expressions.  Parameters are identical to the ones in \mathem{\mmaDef{FMakeP0Rule}}.
		\\[3ex]
		\mathem{\mmaDef{FClearTraceCache}[]} &
		\small Removes all cached trace files from the trace cache directory. \mathem{\mmaDef{FClearTraceCache}[subdirectory]} removes files from a specific subdirectory within the cache.
		\\[3ex]
		\mathem{\mmaDef{FMakeFormMomentumExpansion}[p1,...]} &
		\small Creates a \FORM rule to expand out any scalar products in expressions. The optional momenta arguments specify which momenta to expand.
		Can be passed as a postRepl or preRepl rule to \mathem{\mmaDef{FormTrace}} or \mathem{\mmaDef{FFormSimplify}}. \textit{Warning:} usually, this degrades the performance of \FORM.
		\\[3ex]
		\mathem{\mmaDef{FMakeFiniteTFormMomentumExpansion}[}
		\newline\phantom{.}\hspace{4.42cm}\mathem{p1,...]} &
		\small Creates a \FORM rule to expand scalar products into spatial and temporal parts, given the momenta to expand.
		Separates d-dimensional momenta into (d-1)-dimensional spatial parts and time components. \textit{Warning:} usually, this degrades the performance of \FORM.
		\\[3ex]
		\mathem{\mmaDef{FIterativelySum}[list]} &
		\small Efficiently sums large lists of expressions by breaking them into subsets, repeatedly summing and simplifying until only a single expression remains.
		\mathem{\mmaDef{FIterativelySum}[list, finalSize]} returns a list of specified \mathem{finalSize} with equally-sized terms. Uses parallel processing for optimal performance.
		\\[3ex]
		\mathem{\mmaDef{FDiagramSimplify}[expr]} &
		\small Simplifies diagrammatic expressions by collecting terms and optimising their structure using advanced algorithms to identify common subexpressions and factor them efficiently.
	\end{tabular}
\end{tcolorbox}

\begin{tcolorbox}[title=Code output,
	title filled=false,
	colback=pastelblue!5!white,
	colframe=pastelblue,
	fonttitle=\bfseries]
	\begin{tabular}{p{3.8cm} p{12.0cm}}
		\mathem{\mmaDef{JuliaForm}[expr]} & 
		\small Translates the given mathematical expression to Julia code.
		\\[3ex]
		\mathem{\mmaDef{JuliaCode}[expr]} & 
		\small Generates a return statement for Julia code for the given mathematical expression.
		\\[3ex]
		\mathem{\mmaDef{MakeJuliaFunction}[...]} &
		\small Generates a Julia function definition based on specified options. \newline See \mathem{\mmaDef{Options}[\mmaDef{MakeJuliaFunction}]} for available settings.
		\\[3ex]
		\mathem{\mmaDef{FortranCodeForm}[expr]} &
		\small Translates the given mathematical expression to Fortran code.
		\\[3ex]
		\mathem{\mmaDef{FortranCode}[expr]} &
		\small Generates optimised Fortran code for the given mathematical expression.
		\\[3ex]
		\mathem{\mmaDef{MakeFortranFunction}[...]} &
		\small Generates a Fortran function definition based on specified options. \newline See \mathem{\mmaDef{Options}[\mmaDef{MakeFortranFunction}]} for available settings.
		\\[3ex]
		\mathem{\mmaDef{CppForm}[expr]} & 
		\small Translates the given mathematical expression to C++ code.
		\\[3ex]
		\mathem{\mmaDef{CppCode}[expr]} & 
		\small Generates a return statement for C++ code for the given mathematical expression.
		\\[3ex]
		\mathem{\mmaDef{CppCodeFORM}[expr]} & 
		\small Generates a return statement for C++ code using FORM's output functionality for the given mathematical expression.
		\\[3ex]
		\mathem{\mmaDef{MakeCppClass}[...]} & 
		\small Generates code for a C++ class. \newline The output is customised through its options, see \mathem{\mmaDef{Options}[\mmaDef{MakeCppClass}]}.
		\\[3ex]
		\mathem{\mmaDef{MakeCppHeader}[...]} & 
		\small Generates code for a C++ header. \newline The output is customised through its options, see \mathem{\mmaDef{Options}[\mmaDef{MakeCppHeader}]}.
		\\[3ex]
		\mathem{\mmaDef{MakeCppBlock}[...]} & 
		\small Generates code for a C++ namespace. \newline The output is customised through its options, see \mathem{\mmaDef{Options}[\mmaDef{MakeCppBlock}]}.
		\\[3ex]
		\mathem{\mmaDef{MakeCppFunction}[...]} & 
		\small Generates code for a C++ function/method. \newline The output is customised through its options, see \mathem{\mmaDef{Options}[\mmaDef{MakeCppFunction}]}.
		\\[3ex]
		\mathem{\mmaDef{FUseCppPowr}[bool]} & 
		\small Controls whether to use the \texttt{powr} function for power operations in C++ code generation.
		\\[3ex]
		\mathem{\mmaDef{FormatCppCode}[string]} & 
		\small Uses \texttt{clang-format}, if available on the system, to automatically format a given string of C++ code.
		\\[3ex]
	\end{tabular}
\end{tcolorbox}

\begin{tcolorbox}[title=Global Options,
	title filled=false,
	colback=pastelblue!5!white,
	colframe=pastelblue,
	fonttitle=\bfseries]
	\begin{tabular}{p{5.2cm} p{10.0cm}}		
		\mathem{\mmaDef{FSetGlobalSetup}[setup]} & 
		\small Sets a global setup that is used by all FEDeriK functions when no setup is explicitly provided. 
		This allows for calling functions like \mathem{\mmaDef{FTakeDerivatives}[expr,derivativeList]} without passing the setup each time. \mathem{\mmaDef{FSetGlobalSetup}[]} clears the global setup.
		\\[3ex]
		\mathem{\mmaDef{FSetAutoBuildSymmetryList}[bool]} & 
		\small Sets whether a symmetry list should be automatically built when taking derivatives. Default is True.
		\\[3ex]
		\mathem{\mmaDef{FSetAutoSimplify}[bool]} & 
		\small Sets whether automatic simplification should be applied when taking derivatives and truncating. Default is True.
		\\[3ex]
		\mathem{\mmaDef{FSetDebugLevel}[int]} & 
		\small Sets the amount of debug output given by \FunKit. 
		Default is 0, can be set up to 7. \textbf{Warning:} high values massively degrade performance.
		\\[3ex]
		\mathem{\mmaDef{FSetAlwaysExpandLorentzTensors}[bool]} & 
		\small Specifies whether to set or unset the option \mathem{\mmaDef{ExpandLorentzStructures}} for \FormTracer. Default is \mathem{\mmaDef{True}}.
		\\[3ex]
		\mathem{\mmaDef{FSetCacheDirectory}[folder]} & 
		\small Changes the directory where traced expressions are cached.
		\mathem{\mmaDef{FSetCacheDirectory}[]} resets the cache directory to the default \texttt{/tmp/TraceCache/}.
		\\[3ex]
		\mathem{\mmaDef{FSetRegisterSize}[int]} &
		\small Sets the number of available registers for optimisation in C++ code generation. This is in particular important for calculations on the GPU, where the number of registers is very limited.	The default value is 32, but a larger value on CPU may improve performance.
		\\[3ex]
		\mathem{\mmaDef{FSetNotationA}[]} &
		\small Activates the default two-list object notation: \mathem{obj[\{f1,f2,...\},\{i1,i2,...\}]}. This is the standard \FunKit notation and is active by default at load time.
		\\[3ex]
		\mathem{\mmaDef{FSetNotationB}[]} &
		\small Activates the field[index] object notation: \mathem{obj[f1[i1],f2[i2],...]}. In this notation each argument to a correlation function directly pairs its field with its index.
		\\[3ex]
		\mathem{\mmaDef{FSetCodeOptimization}[bool]} &
		\small Enables (\mathem{\mmaDef{True}}, default) or disables (\mathem{\mmaDef{False}}) the C++ code optimisation pipeline.
		\\[3ex]
		\mathem{\mmaDef{FSetFastMath}[bool]} &
		\small Enables CUDA fast-math intrinsics (\texttt{\_\_expf}, \texttt{\_\_logf}, etc.). Single precision only.
		\\[3ex]
		\mathem{\mmaDef{FSetMaxKernelTerms}[int]} &
		\small Sets max terms per sub-kernel before splitting. Default 500.
		\\[3ex]
		\mathem{\mmaDef{FSetCodePrecision}[precision]} &
		\small Sets code precision. Accepts \mathem{"single"} or \mathem{"double"}. Default \mathem{"double"}.
	\end{tabular}
\end{tcolorbox}

\endgroup

%%%%%%%%%%%%%%%%%%%%
\bibliographystyle{elsarticle-num}
\bibliography{bibliography}
%%%%%%%%%%%%%%%%%%%%

\end{document}